\documentclass[fleqn,usenatbib]{mnras}
\usepackage{xcolor}

\usepackage{newtxtext,newtxmath}
\usepackage{hyperref}
\usepackage[T1]{fontenc}

\DeclareRobustCommand{\VAN}[3]{#2}
\let\VANthebibliography\thebibliography
\def\thebibliography{\DeclareRobustCommand{\VAN}[3]{##3}\VANthebibliography}

\usepackage{graphicx}	% Including figure files
\usepackage{amsmath}	% Advanced maths commands
\graphicspath{{figs/}}
\title[Identifying Dark Star Clusters]{Mass segregation and velocity dispersion as evidence for a dark star cluster}

\author[W. Wu et al.]{
Wenjie Wu,$^{1}$\thanks{ wu.wj.astro@icloud.com }
Pavel Kroupa,$^{1,2}$\thanks{pkroupa@uni-bonn.de}
and Jan Pflamm-Altenburg,$^{1}$\thanks{jpa@hiskp.uni-bonn.de}
\\
$^{1}$Helmholtz-Institut für Strahlen- und Kernphysik (HISKP),
Universität Bonn,
Nussallee 14–16, 
D-53115 Bonn, Germany\\
$^{2}$Charles University in
Prague, Faculty of Mathematics and Physics, Astronomical Institute, v Holesovickach 2, 18000 Praha 8, Czech Republic
}

\date{Accepted 2024 April 26. Received 2024 April 26; in original form 2023 July 25}

\pubyear{2023}

\begin{document}
\label{firstpage}
\pagerange{\pageref{firstpage}--\pageref{lastpage}}
\maketitle

\begin{abstract}
A dark star cluster (DSC) is a system in which the cluster potential is dominated by stellar remnants, such as black holes and neutron stars having larger masses than the long-lived low-mass stars. Due to mass segregation, these remnants are located in the central region of the cluster and form a dark core. We expect that at a few kpc from the Galactic centre, the efficient evaporation of the lower-mass stars caused by the strong tidal force exposes the dark core, because the dynamical properties of the DSC are dominated by the remnants.
Due to the invisibility of the remnants, finding a DSC by observation is challenging. In this project, we use $N$-body simulations to obtain models of DSCs and try to discern observables that signify a DSC. We consider four observables: the mass spectrum, the observational mass density profile, the observational velocity dispersion profile and the mass segregation. The models show that a DSC typically exhibits two distinct characteristics: for a given mass in stars and a given half-light radius the expected velocity dispersion is underestimated when only visible stars are considered, and there is a lack of measurable mass segregation among the stars. These properties can be helpful for finding DSCs in observational data, such as the Gaia catalogue.
\end{abstract}


%%%%%%%%%%%%%%%%%%%%%%%%%%%%%%%%%%%%%%%%%%%%%%%%%%

%%%%%%%%%%%%%%%%% BODY OF PAPER %%%%%%%%%%%%%%%%%%

\section{Introduction}
\label{sec:introduction}
Massive stars have relatively short lifetimes, typically lasting only a few million years. They eventually transform into compact remnants as black holes (BH) or neutron stars (NS), after departing from the main sequence (MS). These remnants possess significantly higher masses compared to their long-lived ($>$ few Gyr), low-mass counterparts. The phenomenon known as the Spitzer instability, as discussed by \citet{spitzer}, leads to their tendency to segregate towards the central region of the star cluster, thereby increasing their population density there. Consequently, black holes cannot maintain energy equipartition with the surrounding main sequence stars. This results in the formation of a highly concentrated and self-gravitating sub-cluster at the core, where the dominant population consists of black holes and neutron stars. This sub-system is referred to as a \textit{dark core} \citep{DSC}. The strong tidal field leads to rapid tidal stripping, exposing the dark core within a few kpc from the Galactic centre. A star cluster in this particular phase is termed a \textit{dark star cluster} (DSC). \citet{DSC} were the first to introduce and study this type of star cluster by conducting a series of N-body simulations of star clusters within the Galactic potential. If a DSC can be observed, it provides an opportunity to constrain the kick velocity distribution of BHs and NSs and gain deeper insights into the mechanisms behind supernova explosions. The dark core remains dynamically active, primarily due to the formation of binary black hole systems (BH-BH) and their subsequent hardening through super-elastic encounters. Consequently, DSCs play a significant role in the emission of gravitational waves (GWs), as discussed by \citet{2021arXiv210801045T}.

The presence of a DSC is contingent upon the fulfillment of several conditions. Once the dark core is established, three-body encounters can directly eject remnants from the core. As a result, the tidal stripping must exert sufficient strength to eliminate a substantial number of bright stars before the dark core depletes itself. There exists a competition of timescales between the tidal stripping of stars from the cluster and the self-depletion of the dark core, determining the emergence of a dark star cluster. Based on two-component simulations, \cite{2013MNRAS.432.2779B} demonstrate that the evolution of the central BH subsystem is influenced by the energy demands of the entire cluster, and the depletion of this subsystem is regulated by the half-mass relaxation time of the cluster. Additionally, the dark core has the capacity to heat the star cluster, resulting in an expansion of the core radius \citep[][]{2007MNRAS.379L..40M,2008MNRAS.386...65M}. In the case of binaries composed of two remnant components, the emission of GW can drive the decay of their relative orbit. Since GW carries linear momentum, and in accordance with linear momentum conservation, the binary's center of mass experiences a recoil or kick when the binary merges.
Previous works \cite[e.g.][]{2008ApJ...686..829H, bhkick} has shown that this kick velocity can reach of thousands of $\mathrm{km \cdot s^{-1}}$, leading to the ejection of remnants. In addition, BHs and NSs are formed through supernova explosions, this can lead to what is commonly known as 'supernova kick-outs.' In simulations, the kick velocity can be drawn from a Maxwellian distribution with a dispersion $\sigma_{\rm kick}$. In this context, the kicks of BHs can be as fast as those of NSs \citep[e.g. ][]{bhkick1, bhkick2,2013MNRAS.434.1355J}. Consequently, the highest value of $\sigma_{\rm kick}$ for BHs is determined by the best fit of the velocity distribution of the observed NSs, which is approximately 265 $\mathrm{km\cdot s^{-1}}$ \citep[average value being of $\approx 420~\mathrm{km\cdot s^{-1}}$,][]{2005MNRAS.360..974H}. In comparison to the escape velocity of a star cluster, which is approximately 1 $\mathrm{km\cdot s^{-1}}$, this kick velocity removes all remnants, rendering the existence of a DSC impossible. However, if the kick velocity dispersion is comparable to the escape velocity of a star cluster, remnants can remain within the star cluster. \cite{2018A&A...617A..69P} quantify the retention fraction of BHs in star clusters with different initial properties assuming different BH kick distributions.

Detecting DSCs through observation is an intricate task, primarily due to the inherent challenges in observing BHs and NSs. In previous studies such as \cite{2018MNRAS.475L..15G} and \cite{2019A&A...632A...3G}, detached Main Sequence-Black Hole (MS-BH) binary systems have been employed to identify stellar-mass BHs in the Galactic globular cluster NGC 3201. However, such binary systems are relatively rare. In the context of DSCs, remnants constitute a significant portion of the cluster potential. Consequently, for specific observables, stars serve as tracers of the cluster potential, aiding in the identification of a DSC.
Previous research has offered observable signatures and properties for the central BH subsystem. Recent investigations regarding the globular cluster M4, as discussed in \cite{2023MNRAS.522.5740V}, reveal the presence of a dark central mass within the cluster. However, using kinematic analysis alone, it remains challenging to differentiate between a point-like dark mass (e.g., an intermediate-mass black hole) and a compact dark core. Based on $N$-body simulations of realistic star clusters, \cite{2016MNRAS.458.1450W} demonstrate that a BH subsystem results in a reduced central surface brightness but a higher central velocity dispersion. Monte Carlo star cluster simulations conducted by \cite{2018MNRAS.479.4652A} and \cite{2018MNRAS.478.1844A,2019MNRAS.485.5345A} have provided constraints on the observable properties of globular clusters that may indicate the presence of a substantial number of BHs. These constraints have been applied to Galactic globular clusters. Moreover, recent works \citep[e.g.][]{2021ApJ...908..224W,2020ApJ...898..162W,2018ApJ...864...13W,2021MNRAS.508.4385A} reveal that the presence of a BH subsystem reduces the degree of mass segregation in globular clusters.

The primary objective of this project is twofold: i) to elucidate how a dark mass component within a star cluster can be identified, and ii) to confirm that the dark mass component primarily comprises massive stellar remnants. Addressing the first issue necessitates the acquisition of velocity-related data, including the velocity dispersion. In order to address the second question, it is imperative to possess position-related information about the stars for quantifying mass segregation and density distribution. Consequently, the computation of the mass density profile, velocity dispersion profile, mass spectrum, and mass segregation becomes imperative.

This paper is organized as follows: Section \ref{sec:computations} provides a description of the simulation setup. Section \ref{sec:res} presents the simulation results and discussions. In Section \ref{sec:con}, we present our conclusions and propose an observational method for identifying a DSC.
\section{Model and simulations}    \label{sec:computations}
\subsection{\texorpdfstring{$N$}{N}-body simulation}

We simulate the evolution of model star clusters within the Milky Way potential. The initial state of these clusters is generated using the \texttt{McLuster} code \citep{mcluster}. The evolution is calculated using the $N$-body code \texttt{PeTar} \citep{petar}, a symplectic and direct integrator designed for $N$-body systems.
In \texttt{PeTar}, the Hamiltonian of the system is divided into short-range and long-range components. For long-range forces, the code employs the \texttt{FDPS} library \citep{fdps} to construct a particle tree, significantly enhancing computational efficiency. Short-range forces are handled using the \texttt{SDAR} library \citep{sdar}. The library uses the slow-down time-transformed explicit symplectic method, which combines the advantages of the symplectic integrator which conserves the Hamiltonian and angular momentum. The high efficiency of the slow-down method can handle the long-term evolution of hierarchical systems and close encounters. The details of the algorithm can be found in \citet{2020MNRAS.493.3398W}.
In terms of binary and single stellar evolution (BSE/SSE), \texttt{PeTar} relies on the \texttt{BSE} code \citep{2002MNRAS.329..897H,bse}, an analytical and extensively tested stellar evolution model.

We generate all models following a Plummer model \citep{plummer}. The 3-dimensional mass density profile, denoted as $\rho(r)$, and the velocity dispersion profile, represented by $\sigma(r)$, are defined as outlined in e.g. \cite{plummerprop} and \cite{2003gmbp.book.....H}:
\begin{equation}
	\label{eq:plummar}
	\begin{aligned}
		\rho(r) &= \frac{3M_{\rm c}}{4\pi R_{\rm pl}^3}\left[1+\left(\frac{r}{R_{\rm pl}}\right)^2\right]^{-\frac{5}{2}} \\
		\sigma^2(r) &= \frac{GM_{\rm c}}{2R_{\rm pl}}\left[1+\left(\frac{r}{R_{\rm pl}}\right)^2\right]^{-\frac{1}{2}}
	\end{aligned},
\end{equation}
where $M_{\rm C}$ is the total mass of the star cluster and $R_{\rm pl}$ the Plummer radius. In our setup, we use the 3-dimensional half mass radius $R_h=(2^{2/3}-1)^{-1/2}R_{\rm pl}$ (containing 50\% of the total mass) as the scale radius. We assume that the cluster initially possesses an isotropic velocity distribution function, meaning that the 1D tangential and radial components of the velocity dispersion are not distinct. Our models are initialized in virial equilibrium, closely resembling open clusters. As a result, we do not consider the early, violent emergence through gas expulsion from the compact and deeply embedded phase, as outlined in \cite{2001MNRAS.321..699K}, \cite{2017AA...597A..28B} and \cite{2022AA...660A..61D}. Our focus here is on the study of the later dark phase.

For the initial mass functions (IMF), we employ the canonical Kroupa IMF \citep{2001MNRAS.322..231K}. The number of stars with mass in the interval of $m$ and $m+\mathrm{d}m$ (in units of $M_\odot$) is given by $\mathrm{d}N=\xi(m)\mathrm{d}m$, with $\xi(m)$ defined as
\begin{equation}
	\xi(m) = k\left\{
	\begin{aligned}
		2m^{-\alpha_1}, &\quad 0.08~ M_\odot \leq m < 0.5~ M_\odot \\
		m^{-\alpha_2}, &\quad 0.5~ M_\odot \leq m < 1 ~M_\odot\\
		m^{-\alpha_3}, &\quad 1~ M_\odot \leq m_{\rm max}(M_{\rm c}) \leq m_{\rm max *}
	\end{aligned}
	\right. ,
\end{equation} 
where $\alpha_1 = 1.3$, $\alpha_2=\alpha_3 = 2.3$. $k$ is the normalization constant. Note, the factor 2 for $0.08~ M_\odot \leq m < 0.5~ M_\odot$ is valid only for $\alpha_2-\alpha_1=1$
The physical upper limit of stellar masses is $m_{\rm max *} = 150~ M\odot$, as indicated in e.g. \cite{maxmass}. $m_{\rm max}(M_{\rm c})$ represents the maximal stellar mass within the cluster and is dependent on the total mass of the cluster \citep[see][]{maxclutermass1, maxclustermass2, 2023A&A...670A.151Y}:
\begin{equation}
\label{eq:mmax}
	m_{\rm max}(M_{\rm c}) =  a\cdot M_{\rm c}^{b}.
\end{equation}
For our models that have $M_{\rm c} > 3300 ~M_\odot$, $a = 2.05$ and $b=0.36$, as obtained by fitting eq. (\ref{eq:mmax}) to the observational data containing clusters which are more massive than 3300 $M_\odot$.

In addition, we include two models with primordial mass segregation to explore potential effects following the method described by \cite{2008ApJ...685..247B} for initializing a mass-segregated cluster in virial equilibrium \citep[see also ][for an alternative approch]{2008MNRAS.385.1673S}{}{}. As observed in \citet{2008ApJ...685..247B} and \citet{ms2}, there is evidence of primordial mass segregation in globular and open clusters.
This setup is implemented by setting the parameter $S=1$ in \texttt{McLuster}, with $S=0$ defining a cluster without mass segregation. In brief, for the case $S=1$ the code initially generates stellar orbits according to the Plummer model and orders them based on specific energy. Subsequently, the code arranges the generated masses of stars and pairs both the orbital array and mass array together.

In terms of stellar evolution, the determination of remnant mass is based on the rapid supernovae scenario as outlined in \citet{Fryer2012ApJ...749...91F}. The calculations also account for pair-instability \citep{Belczynski2016A&A...594A..97B} and electron capture supernova \citep{Belczynski2008ApJS..174..223B}. A solar metallicity of $Z=0.02$ \citep{von2016ApJ...816...13V} is assumed for all models.
To retain all remnants within the cluster, we introduce no natal kicks for NSs and BHs. However, we also conducted a model incorporating natal kicks and qualitatively discuss the results. It should be noted that remnants can still escape from the cluster due to interactions with other stars or remnants. The  parameters used for the BSE/SSE routine in \texttt{PeTar} are listed in Appendix \ref{sec:bseparameter} tab. \ref{tab:petarbseparameter}.
We computed the evolution of 6 star cluster models in this study. Tab. \ref{tab:models} presents a list of the models used in this project. We selected a Galactocentric distance of 2 kpc for our simulations, as this distance allows the DSC phase to occur relatively early, resulting in shorter simulation runtimes.
models 1 and 2 incorporate primordial mass segregation. For model 1, we adopted an initial half-mass radius of 2.5 pc. Model 2 was used for the purpose of comparing the influence of the initial half-mass radius, and thus, an initial half-mass radius of 1.5 pc was employed.
Models 3 and 4 utilized the same setups as models 1 and 2, respectively, but did not incorporate primordial mass segregation. Model 5 introduced natal kicks. We use a 18\% initial binary fraction in model 6. The binary fraction is defined by
\begin{equation}
    f_{\rm b} = \frac{N_{\rm b}}{N_{\rm s} + N_{\rm b}},
\end{equation}
where $N_{\rm b}$ is the number of binary systems and $N_{\rm s}$ is the number of single stars. \texttt{McLuster} randomly selects which star is in a binary system. For stars with mass a $> 5$ $M_\odot$, we employ the \cite{2012Sci...337..444S} distribution, which is derived from O star samples of six nearby Galactic open clusters. The distribution functions of initial periode, eccentricities, mass ratio and semi-major axis length are based on \cite{2015ApJ...805...92O} and \cite{2017MNRAS.471.2812B}. 
It is important to note that general relativity effects were not considered in these models.

\begin{table*}
	\caption{List of used models, where $M_{\mathrm{c}, 0}$ is the initial cluster's mass, $R_{\rm h,0}$ the 3-dimensional half mass radius, $R_G$ the Galactocentric distance, $\rho_0$, the central density, $S$ the mass segregation parameter and $f_{\rm b}^{\rm init}$ the initial binary fraction. The penultimate column represents the age of each model at the end of the simulation. The last column shows the natal-kick choice.}
	\begin{tabular}{lccccccccr}
		\hline
		model & $M_{\mathrm{c}, 0}$& $R_{\rm h,0}$& $R_G$ & $\rho_0$ & $S$ & $f_{\rm b}^{\rm init}$& evolved time  & natal kick\\ 
		 &  [$M_\odot]$ & [pc] & [kpc] & [$M_\odot\mathrm{\cdot pc^{-3}}$] & & & [Myr]& &\\
		\hline
		1 & $3\times 10^4$ & 2.5 & 2.0 & 11158 & 1 & 0\%& 3722& no \\
		2 & $3\times 10^4$ & 1.5 & 2.0 & 51660 & 1 & 0\%& 4000&no \\
            3 & $3\times 10^4$ & 2.5 & 2.0 & 11158 & 0 & 0\%&6000&no \\
		4 & $3\times 10^4$ & 1.5 & 2.0 & 51660 & 0 & 0\%&8000&no \\
            5 & $3\times 10^4$ & 2.5 & 2.0 & 11158 & 0 & 0\%&6000&yes \\
		6 & $3\times 10^4$ & 2.5 & 2.0 & 11158 & 0 & 18\%&4000&no \\
		\hline
	\end{tabular}
	\label{tab:models}
\end{table*}
We utilize the Milky-Way potential as the external tidal field, and this is implemented through the \texttt{galpy} library \citep{galpy}. The library is integrated into \texttt{PeTar}. This potential is the result of combining potentials for a spherical bulge (power-law with a cut-off), an effective dark matter halo (following the Navarro–Frenk–White profile), and a Miyamoto-Nagai disk component.
The potential is scaled in such a way that the Sun orbits at $R_G=8$ kpc with a circular velocity of 220 $\mathrm{km\cdot s^{-1}}$. Both models assume circular orbits around the Galactic centre.
In order to expedite the computation, a particle is eliminated from the simulation when it crosses a fixed radius of $r_{\rm remove} = 80$ pc from the cluster centre (specified to be the center-of-mass position). In the subsequent discussions within this paper, we exclusively consider stars and remnants located within the tidal radius, as in \cite{clusterintidal}:
\begin{equation}\label{eq:rt}
	R_{\rm t} = \left(\frac{M_{\rm c}}{3M_{\rm G}}\right)^{1/3}R_{\rm G},
\end{equation}
where $M_{\rm G}$ the mass of the effective point mass galaxy and $R_{\rm G}$ is the distance between the cluster and the Galactic centre. Here $R_{\rm G} = 2$ kpc and $M_{\rm G}$ is estimated by the central external potential $\Phi_{\rm ext}$ of the cluster i.e. $M_{\rm G} = -\Phi_{\rm ext}R_{\rm G}\cdot G^{-1}$. To determine $M_{\rm c}$ and $R_{\rm t}$ at each time step, we initially calculate the density centre of the star cluster based on the local density of each star \citep[see][]{1985ApJ...298...80C}. This is estimated from the mass within a sphere containing the six nearest particles, and the density centre is defined as the density-weighted mean position of all stars. We then solve equation \ref{eq:rt}, with $M_{\rm c} = M(<R_{\rm t})$. Since the cluster continually loses stars and remnants during its evolution, the tidal radius consistently decreases. The value of $r_{\rm remove}$ is approximately 8 times larger than the initial tidal radius, which is approximately 10 pc. Consequently, we can retain all pertinent information throughout the simulation. Note: In tab. 1, the "evolved time" refers to the duration of the simulation. For models 1, 2, and 3, due to their shorter lifetimes, our analysis of the results was halted when only 20 particles remained in the cluster.

The calculations are done on a desktop computer with an 8-core \texttt{AMD 3700x} CPU and a \texttt{Nvidia 1660 Ti} GPU and with a calculation time of about 200 hours.

\section{Results} 
\label{sec:res}
In this section, we begin by examining the properties of a DSC through the use of model 1 and model 2.  Our initial step involves identifying a suitable definition for the DSC phase and observing the structural evolution. We then delve into a discussion of the projected properties during the DSC phase, as well as the presence of mass segregation.
Subsequently, we explore the effects of initial mass segregation, initial half-mass radius, primordial binaries, and natal kicks by employing all models.

\subsection{Definition of a DSC}
A dark star cluster is defined to be an apparently  unbound system. \cite{DSC} defined the DSC phase by the virial ratio,
\begin{equation}
	\label{eq:Q}
	Q = -\frac{E_{\rm kin}}{E_{\rm pot}},
\end{equation}
where $E_{\rm kin}$ is the total kinetic energy and $E_{\rm pot}$ is the total potential energy of all constituents within $R_{\rm t}$. The initiation of the DSC phase occurs when the observed Q-value reaches 1.0, i.e. the cluster becomes an unbound system. However, this criterion ($Q>1$) can lead to potential confusion. For instance, a binary system with one visible star component and one invisible remnant component can contribute a significant amount of kinetic energy. In further analysis, if a pair of particles has a negative two-body binding energy and their separation is less than 0.1 pc, we treat them as a binary system. When calculating the Q-value, a binary system is regarded as a single particle, and we use the center-of-mass (c.m.) position and velocity as the position and velocity of this composite particle. The total mass of the binary system is taken as its mass. 
If a binary consists of one luminous star and one dark component (BH, NS, or WD), they will be excluded when extracting observational properties. This exclusion is necessary because, in observations, high-velocity stars are often considered as field star contamination.
In this project, we employ two additional conditions to define the DSC phase:
\begin{itemize}
    \item The remnants must exert a greater influence on the cluster potential than the stars, meaning that the ratio of the true potential energy to the observed potential energy should exceed 2.
    \item When considering all stellar types together with the remnants within the cluster, the Q-value should be less than 1, signifying that the cluster may appear to be unbound but is not in actuality.
\end{itemize}

\begin{figure*}
    \centering
    \includegraphics[width=\textwidth]{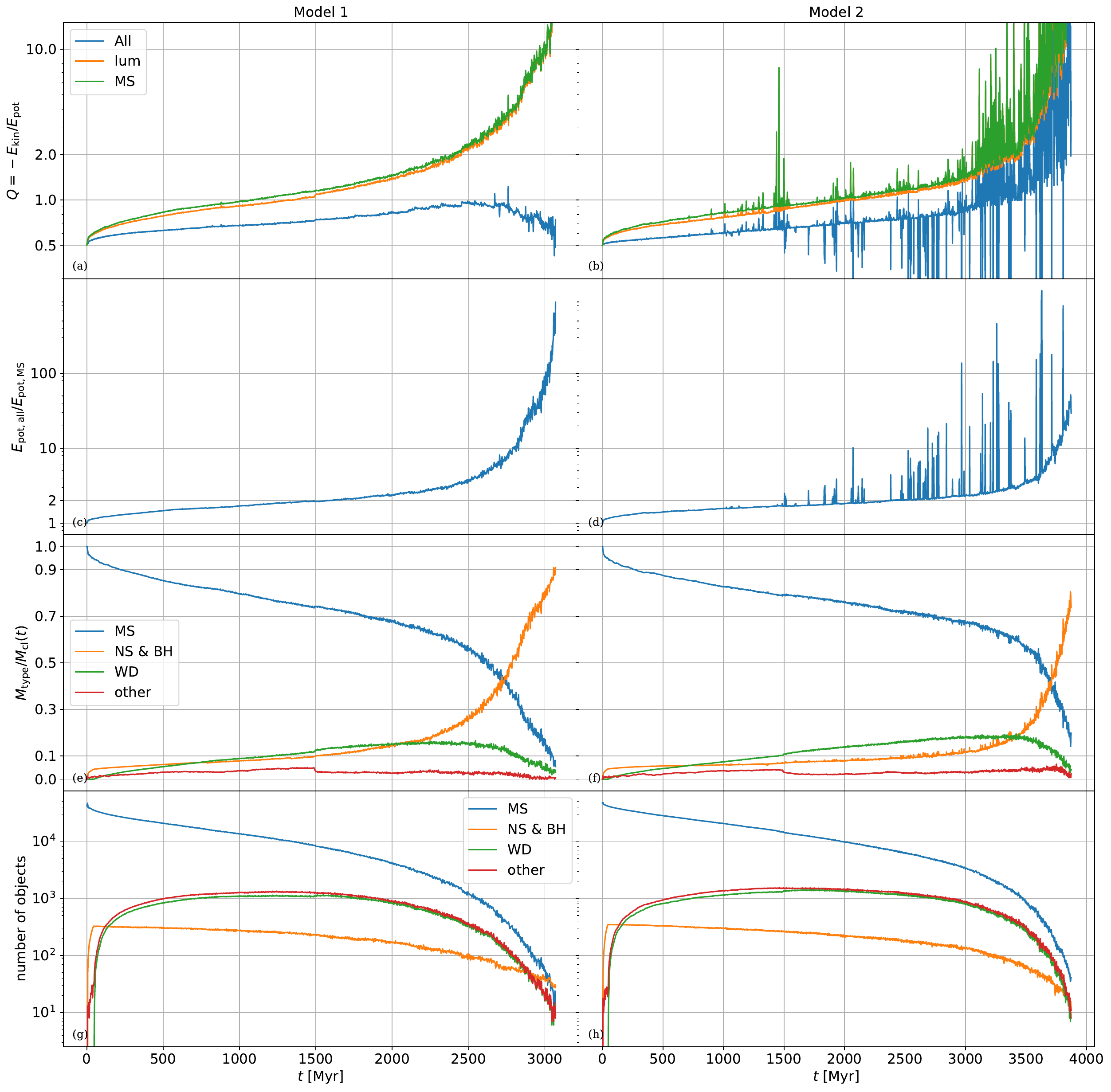}
    \caption{The evolution of the Q-value is depicted in panel a and b, while the ratio of potential energy is shown in panel c and d. Additionally, the ratio of the mass of each component to the total mass is presented in panel e and f, and the number of objects is displayed in panel g and h for model 1 and 2. In panel a and b, the label "All" signifies that all stellar types including remnants are taken into account, "lum" denotes the consideration of only stars, including MS stars, giants and naked helium stars, 'MS' indicates that only MS stars are considered. In panel e, f, g, and h, 'NS \& BH' represents the consideration of only NSs and BHs, 'WD' implies the consideration of only WDs, and 'other' includes all other stellar types, such as giants and naked helium stars. The thick tick on the $x$-axis marks the time.}
    \label{fig:fig112}
\end{figure*}

We present the $Q$-value alongside the potential ratio, mass fraction of each component, and the count of each component in Fig. \ref{fig:fig112}, which illustrates the results for model 1 and 2. In addition to MS stars, NSs and BHs, we also include the number and mass fraction of WDs and other luminous ingredients (giants and naked helium stars). The mass fraction attributed to luminous leftover objects consistently remains below 5\%. 
In the observational context, considering the MS stars alone allows the estimation of a more accurate surface mass density profile from the surface brightness profile, as MS stars follow a straightforward mass-to-light ratio. WDs, on the other hand, are faint and challenging to observe. Given their similar masses to MS stars, they co-evolve with them. Therefore, in subsequent analyses, we exclusively consider MS stars to extract the observational properties of a star cluster. These properties are referred to as 'apparent properties,' while the 'actual properties' can be derived by accounting for all stellar types within the cluster. We use the snapshot at 2000 Myr to conduct further analysis.

In the evolution of the $Q$-value for model 2, noticeable sharp peaks appear as noise (panel b of Fig. \ref{fig:fig112}). This phenomenon is attributed to a small number of stars and remnants briefly crossing the tidal radius and quickly returning. Interestingly, this does not occur in model 1. The primary distinction between these two models lies in their initial half-mass radii. Model 2's higher core density enhances the efficiency of two-body encounters, resulting in the ejection of certain remnants from the core. Additionally, sharp peaks emerge after 1500 Myr in the ratio of potential energy (panel d of Fig. \ref{fig:fig112}). Due to the presence of noise, the initiation of the DSC phase can only be determined through visual inspection of the plots. We identify that at 2700 Myr, the cluster indeed exhibits $Q>1$, and the potential ratio surpasses 2. We use this snapshot to conduct further analysis.

\begin{figure*}
    \centering
    \includegraphics[width=\textwidth]{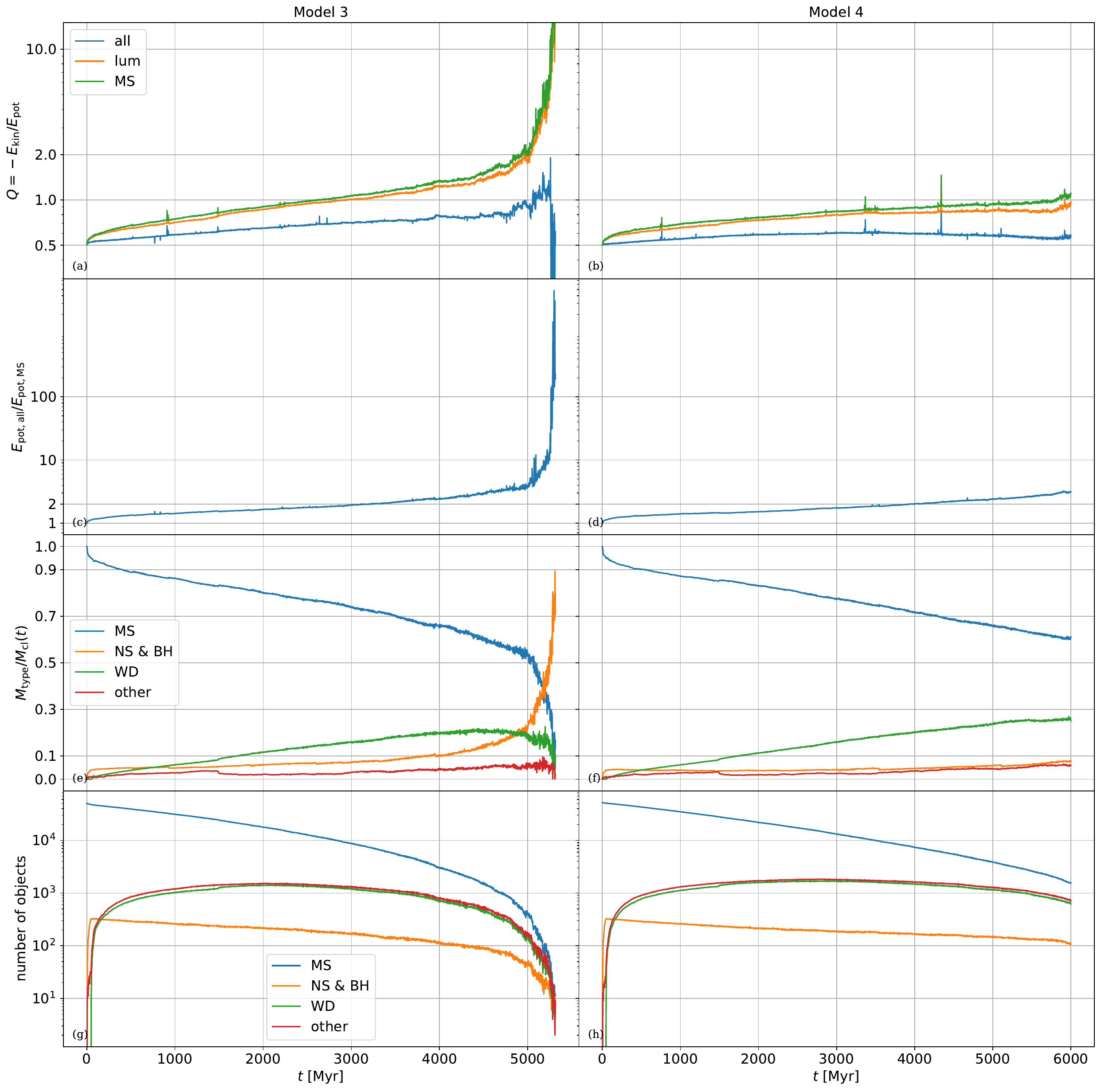}
    \caption{As Fig. \ref{fig:fig112} but for model 3 and 4.}
    \label{fig:fig134}
\end{figure*}

\begin{figure*}
    \centering
    \includegraphics[width=\textwidth]{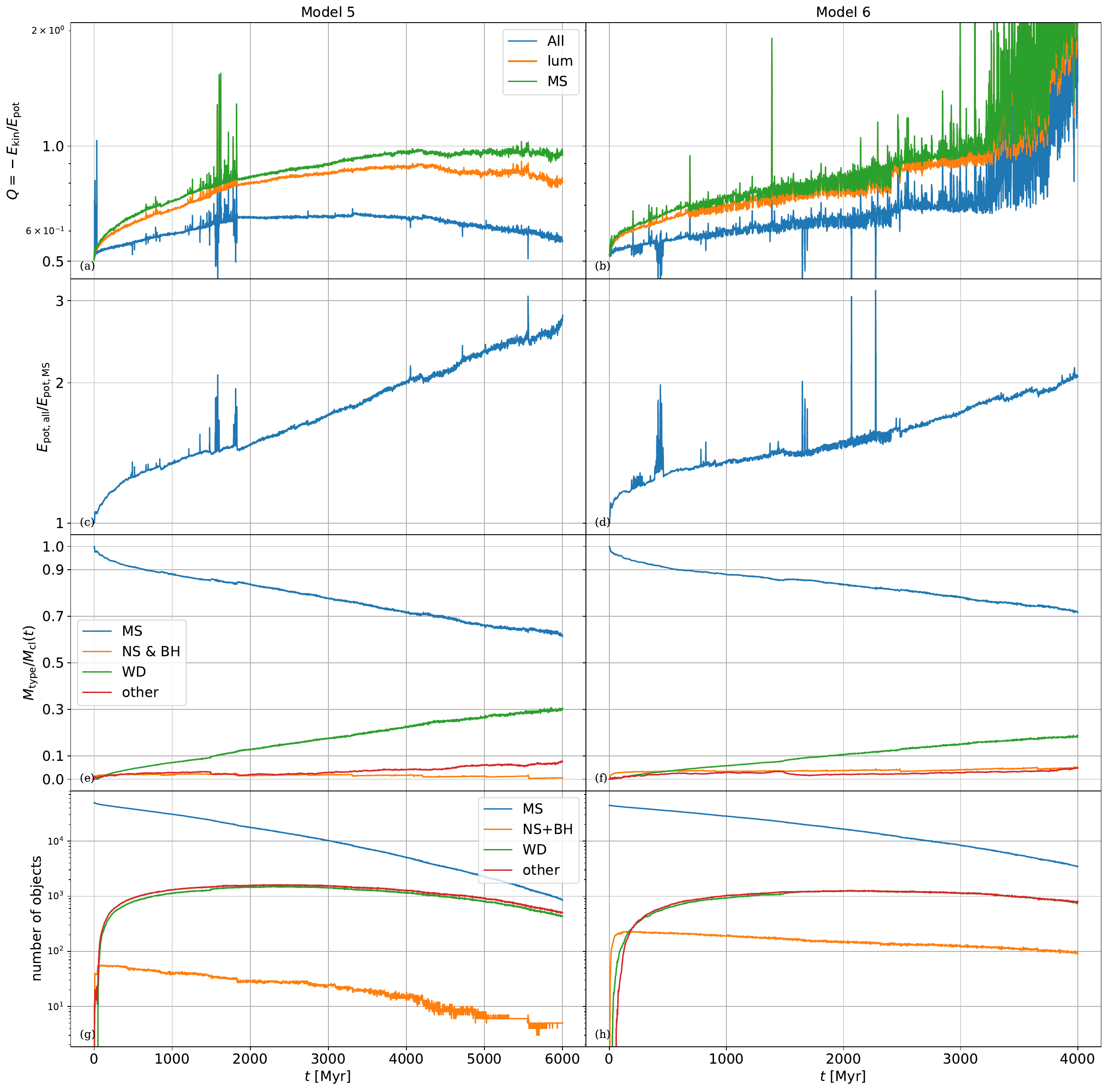}
    \caption{As Fig. \ref{fig:fig112} but for model 5 and 6.}
    \label{fig:fig156}
\end{figure*}

Applying the same methodology, we use the snapshot at 4000 Myr for model 3 (see left panels of Fig. \ref{fig:fig134}). However, for model 4, 5, and 6 (right panels of Fig. \ref{fig:fig134} and Fig. \ref{fig:fig156}), no DSC phase can be discerned. This conclusion is further supported by the analysis of mass fractions (see panels e and f)
The mass fractions of NSs and BHs collectively amount to approximately 10\% of the cluster when they enter the DSC phase. This observation aligns with the findings of \citet{2013MNRAS.432.2779B}, who established that, in a two-component system, the mass fraction of the heavier components should exceed 10\% to achieve a lower evaporation rate of the subsystem compared to the cluster as a whole. 
For model 4, 5, and 6, the mass fractions of NSs and BHs consistently remain below 10\%.

\subsection{Structure evolution}
\label{sec:structevol}
\begin{figure*}
    \centering
    \includegraphics[width=\textwidth]{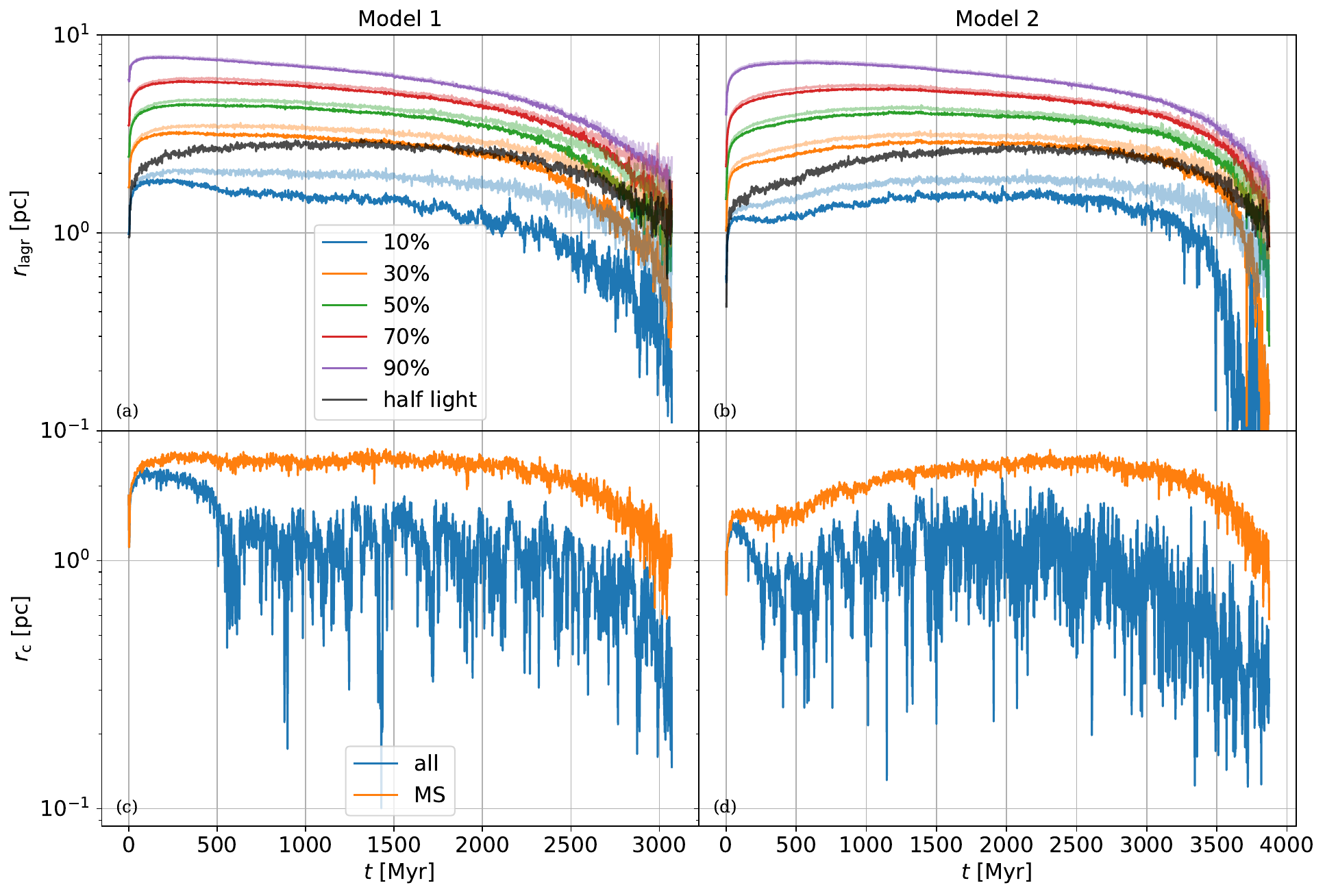}
    \caption{The structural evolution of model 1 and 2 is depicted in the following manner: \textit{panel a and b} The evolution of 10\%, 30\%, 50\% (i.e., the half-mass radius $r_{\rm h}$), 70\%, and 90\% Lagrangian radii. The light blue, light orange, light green, light red and light purple lines in panel a and b correspond to the same Lagrangian radii, but they encompass only MS stars. The black line represents the projected half-light radius, derived only from MS stars. The $x$-axis serves as the line-of-sight. \textit{panel c and d}: The evolution of the core radius for all stellar types (blue lines), solely for MS stars (orange lines).}
    \label{fig:struct12}
\end{figure*}

Fig. \ref{fig:struct12} provides a succinct overview of the structural evolution of model 1 and 2, incorporating Lagrangian radii and core radius. When considering only MS stars, it is observed that the apparent 10\%, 30\%, and 50\% Lagrangian radii are greater than the actual radius (Panel a and b). This observation suggests that remnants are concentrated in the central region of the cluster.
Observationally, the core radius, denoted as $r_{\rm c}$, represents the radius at which the surface luminosity density (or projected surface mass density) drops to half of its central density. In this project, the definition of $r_{\rm c}$ is based on the density-weighted average of the distance of each star from the density center \citep{coreradius}. This definition is commonly employed in theoretical investigations and is correlated with the concentration of the star cluster. A smaller core radius signifies a more concentrated cluster.
\cite{ccosc} demonstrate that the evolution of the core radius undergoes oscillations due to core collapse and binary formation. The formation of binaries causes the core to heat and subsequently expand. However, in a DSC, stars in the core will be pushed out and cannot serve as tracers.
Panel c and d of Fig. \ref{fig:struct12} illustrate the evolution of the core radius. These oscillations only become apparent when all stellar types or NSs and BHs alone are considered. When solely MS stars are taken into account, the core radius is both larger and remains constant following the early expansion phase. It is evident that remnants play a dominant role in the dynamical properties within the central region.
The early expansion of the cluster is primarily instigated by stellar evolution. The significant mass loss resulting from the evolution of massive stars reduces the overall cluster mass, subsequently leading to a decrease in potential energy. In the very late stages, both the core and Lagrangian radii experience a reduction. This reduction is attributed to the loss of stars and remnants.

\begin{figure*}
    \centering
    \includegraphics[width=\textwidth]{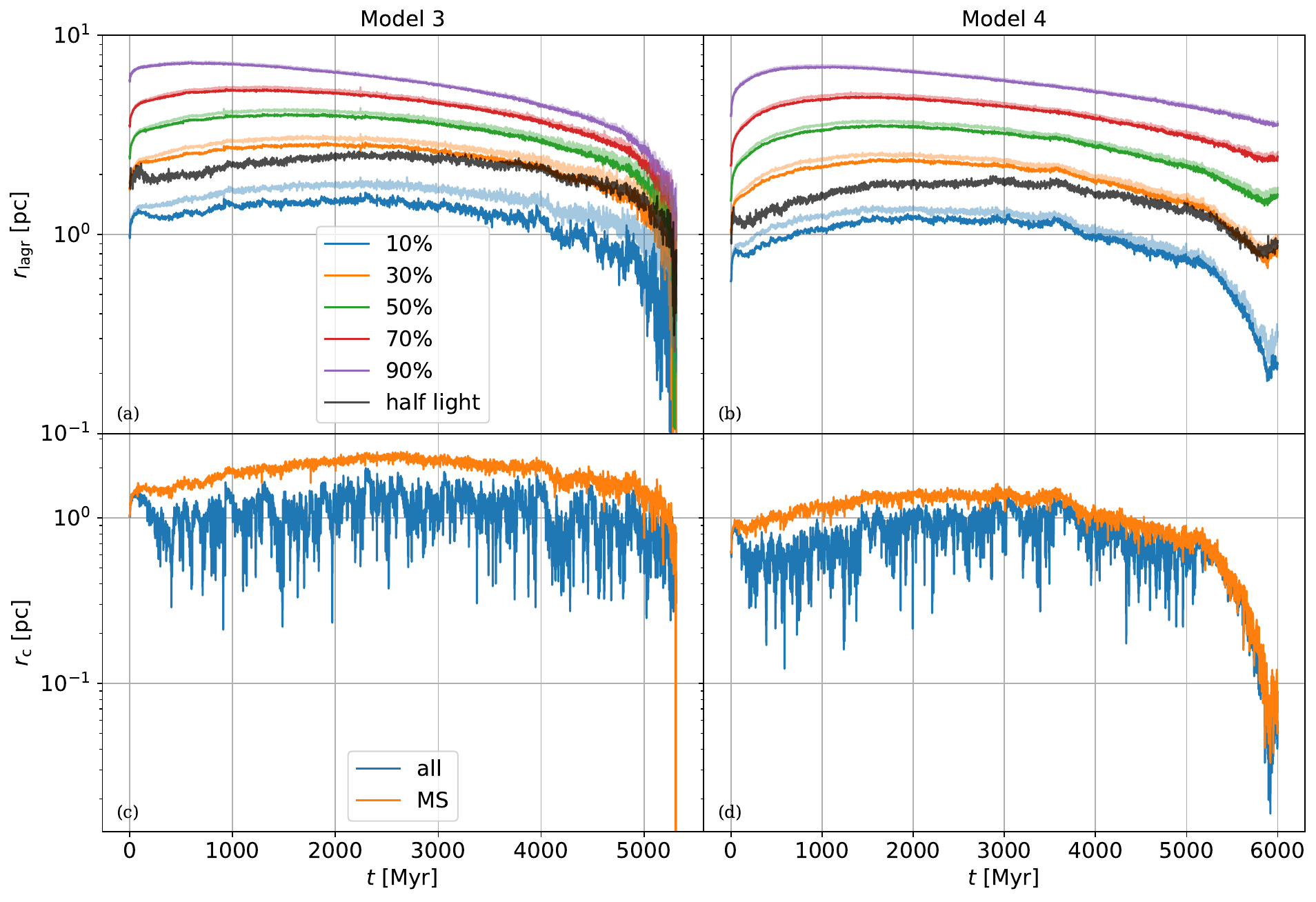}
    \caption{As Fig. \ref{fig:struct12} but for model 3 and 4.}
    \label{fig:struct34}
\end{figure*}

\begin{figure*}
    \centering
    \includegraphics[width=\textwidth]{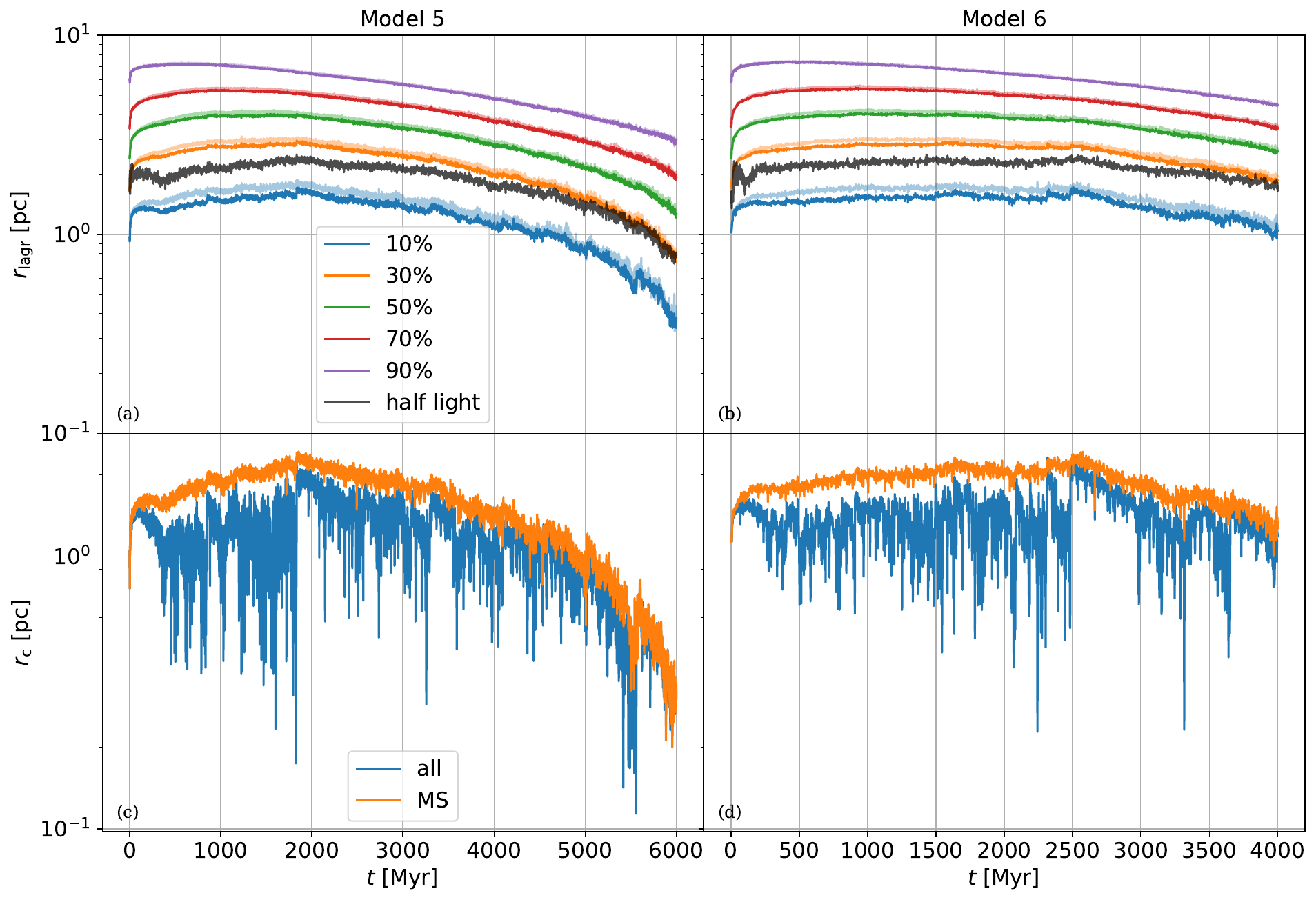}
    \caption{As \ref{fig:struct12} but for model 5 and 6.}
    \label{fig:struct56}
\end{figure*}
\subsection{Observables}
In this subsection, we delve into observables that can be derived from photometry. 
We broadly outline the evolution of the half-light radius alongside the central surface mass and luminosity density. Our analysis extends to examining the surface mass profile and line-of-sight (l.o.s.) velocity dispersion. Subsequently, we gauge the extent of mass segregation within the cluster by employing the minimum-spanning tree algorithm. Additionally, we provide the evolution of the mass spectrum.

\subsubsection{Central surface mass and luminosity}
We selected the $x$-axis to represent the l.o.s. (the galaxy disk is lies on the $x$-$y$ plane) and present the projected half-light radius alongside the Lagrangian radius for each model in panels a and b of Figures \ref{fig:struct12}, \ref{fig:struct34}, and \ref{fig:struct56}. For all models, it is observed that at the later stage, the half-light radius closely aligns with the 30\% Lagrangian radius, calculated solely from MS stars.

The interactions between remnants and stars play a transformative role in altering the cluster's dynamical properties. When considering mass segregation, it is noteworthy that remnants are most likely to interact with the presently most massive stars, which are typically situated near the cluster's core. Despite the fact that the most massive star in the cluster only weighs around 1 $M_\odot$, it remains a low-mass object when compared to NSs and BHs. Consequently, stars tend to be displaced from the core due to the redistribution of energy. This phenomenon results in a less--observable core collapse, as corroborated in Section \ref{sec:structevol}.

\begin{figure*}
    \centering
    \includegraphics[width=\textwidth]{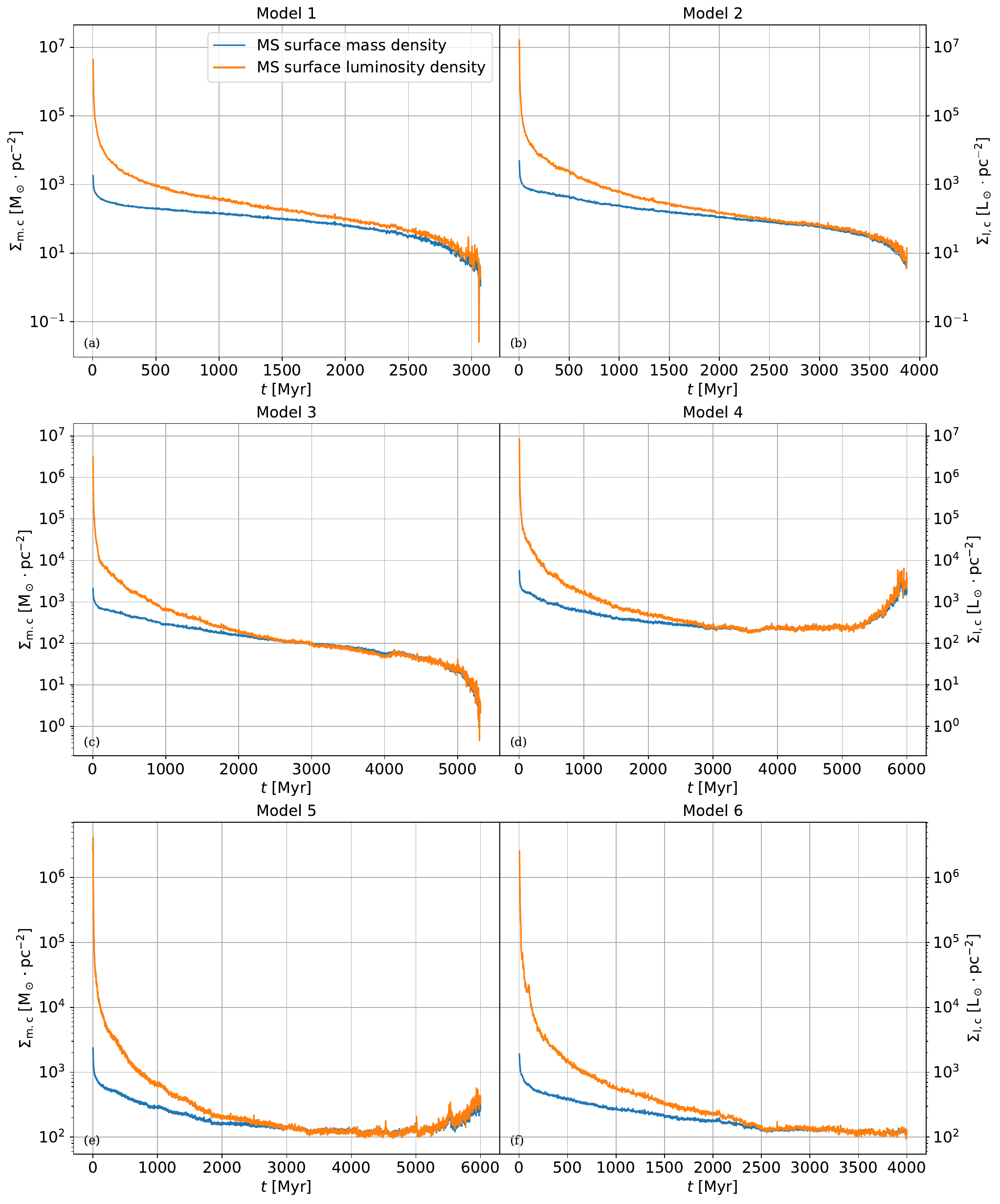}
    \caption{The central surface mass ($\Sigma_{\rm m, c}$) and luminosity density ($\Sigma_{\rm l, c}$) for each model are presented on dual $y$-axes. The central region is delineated within the core radius as depicted in panels c and d of Figures \ref{fig:struct12}, \ref{fig:struct34}, and \ref{fig:struct56}.}
    \label{fig:surfacecentral}
\end{figure*}

In Figure \ref{fig:surfacecentral}, we present an overview of the evolution of the central surface mass and luminosity density for MS stars, utilizing the core radius derived from Section \ref{sec:structevol}. It is observed that for models 1, 2, and 3, the presence of a dark core significantly diminishes the concentration of MS stars in the central region, resulting in a decreased surface mass and luminosity density during the DSC phase. 
\subsubsection{Surface mass density and l.o.s. velocity dispersion}

From an observational perspective, it is often feasible to capture just a single snapshot of the cluster. In the event of core collapse, this leads to the manifestation of a central cusp within the mass density profile. To address this, we utilize the 2D cumulative mass profile $M_{\rm 2D}(<R)$ and fit 
 a Plummer model, with $R$ representing the projected radial distance to the density center. This approach aligns with the method employed in \cite{2015ApJ...800....9M}, as the cumulative profile mitigates statistical uncertainties that arise due to data binning. By integrating equation \ref{eq:plummar} along the l.o.s., the surface mass density profile is obtained,
\begin{equation}
    \Sigma(R) = \frac{M_{\rm c}R_{\rm pl}^2}{\pi \left(R_{\rm pl}^2+R^2\right)^2},
\end{equation}
The cumulative profile i.e. the function to be fitted is
\begin{equation}
    M_{\rm 2D}(<R) = 2\pi\int_{0}^{R}\Sigma(R^\prime)R^\prime\mathrm{d}R^\prime = \frac{M_{\rm c}R^2}{R_{\rm pl}^2+R^2}.
\end{equation}
Note, the projected half mass radius, $r_{\rm h, 2D}$, is equal to the Plummer radius $R_{\rm pl}$. By assuming isotropy and applying the Jeans equation, the l.o.s. velocity dispersion $\sigma_{\rm los}\left(R\right)$ profile follows \citep[][]{2003gmbp.book.....H}
\begin{equation}
\begin{aligned}
    \sigma^2_{\rm los}\left(R\right) &= \frac{2}{\Sigma\left(R\right)}\int_{R}^{\infty}\frac{r\rho\left(r\right)\sigma_r^2\left(r\right)}{\sqrt{r^2-R^2}}\mathrm{d}r \\
    & = \frac{3\pi}{64}\frac{GM_{\rm c}}{R_{\rm pl}}\left( 1+\frac{R^2}{R^2_{\rm pl}} \right)^{-\frac{1}{2}}
\end{aligned}
\end{equation}
where $\rho\left(r\right)$ and $\sigma_r^2\left(r\right)$ follow the first and
the second line in equation \ref{eq:plummar}, respectively. 

\begin{figure*}
	\centering 
	\includegraphics[width=\textwidth]{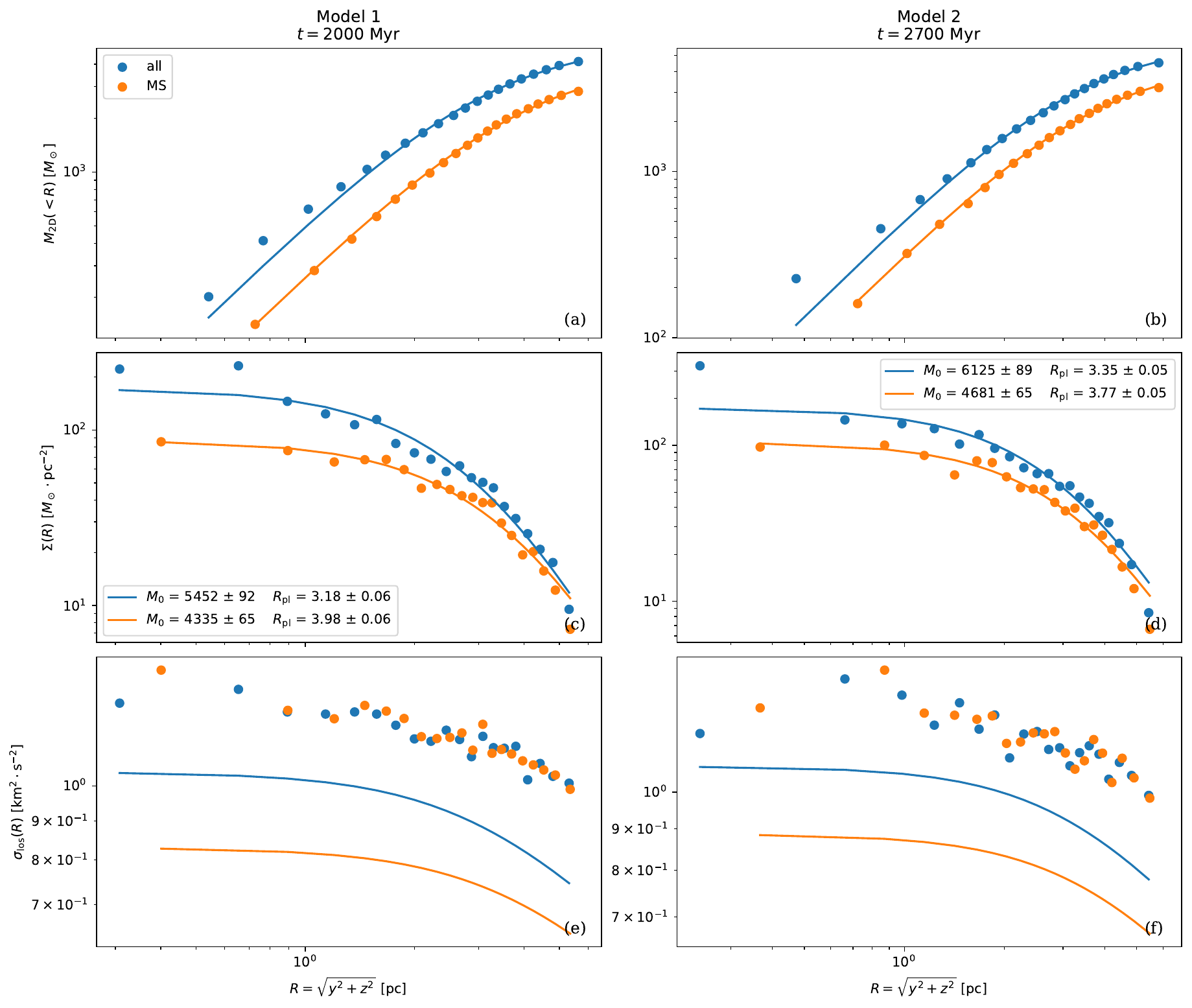}
	\caption{The cumulative surface mass profile (panel a and b), the surface mass density profile (panel c and d), and the l.o.s. velocity dispersion profile (panel e and f) are depicted for model 1 and 2. The time point selected corresponds to the DSC phase for both clusters (see Fig. \ref{fig:fig112}). We adopt the $x$-axis as the l.o.s., with each bin encompassing 5\% of the total mass. Blue dots represent data that was obtained using all stellar types and orange dots represent measurements only when MS stars are used. Plummer fitting is executed utilizing the cumulative surface mass profile in panel a and b, the blue and orange line represent the fitting result. The results of the fitting for the two parameters, $M_{\rm c}$ and $R_{\rm pl}$, are presented in panel c and d. In panel e and f, blue and orange lines are not regression for the l.o.s. velocity dispersion profile, but instead, the profiles are compared with predictions derived from eq. (\ref{eq:plummar}).}
	\label{fig:surface12} 
\end{figure*}

In Fig. \ref{fig:surface12}, we present the fitting results for models 1 and 2. Panel a and b present the measured cumulative surface mass profile for MS stars exclusively and for all stellar types collectively. Panel c and d showcase the projected surface mass density profile along with the corresponding fitting results for both cases. Our observations reveal that the actual profiles exhibit an over-density at the center. However, when solely considering MS stars, the density profiles exhibit no discernible evidence of core collapse and closely adhere to the Plummer model.

\begin{figure*}
	\centering 
	\includegraphics[width=\textwidth]{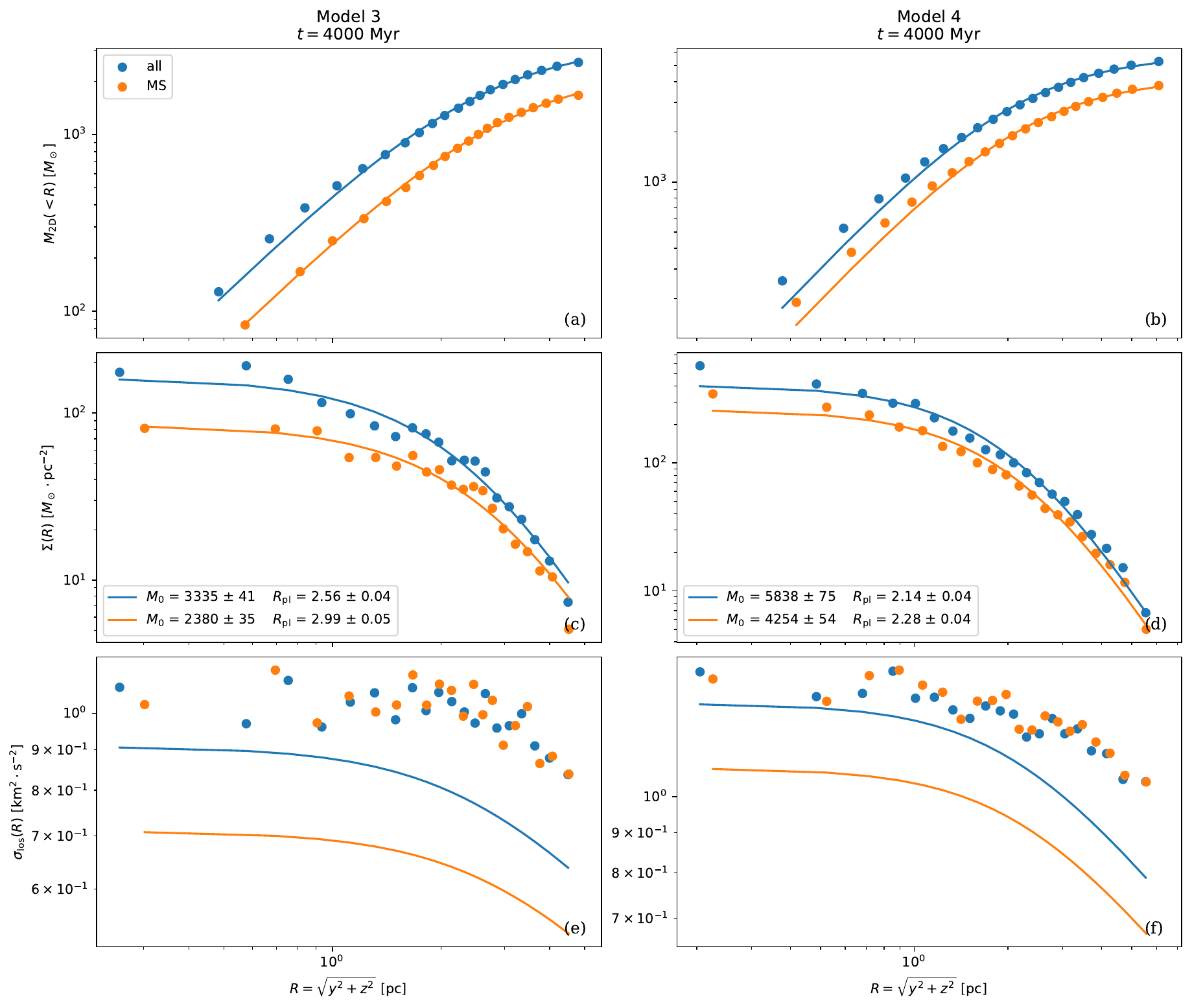}
	\caption{As Fig. \ref{fig:surface12} but for model 3 and 4.}
	\label{fig:surface34} 
\end{figure*}

\begin{figure*}
	\centering 
	\includegraphics[width=\textwidth]{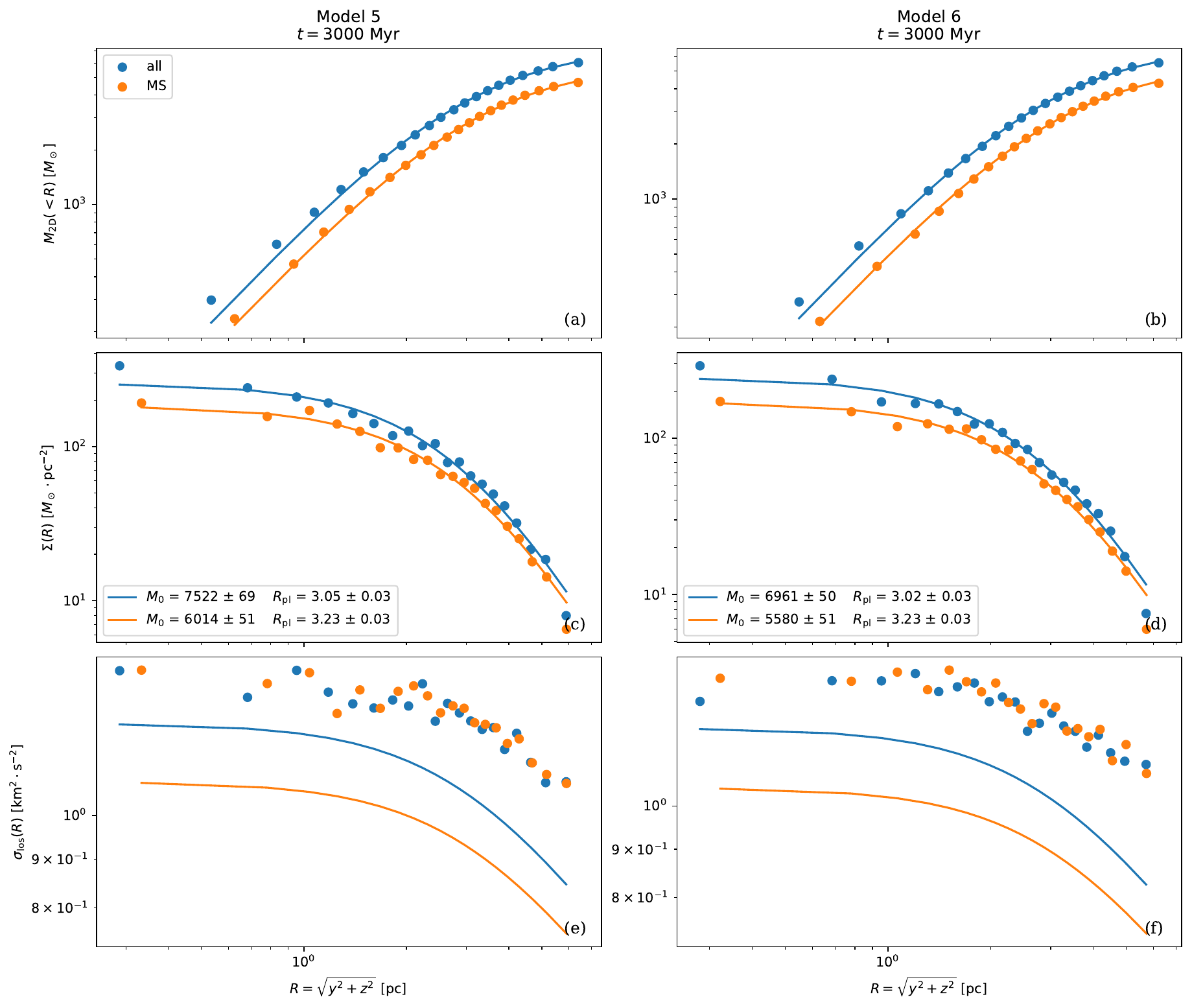}
	\caption{As Fig. \ref{fig:surface12} but for model 5 and 6.}
	\label{fig:surface56} 
\end{figure*}

If we assume that MS stars provide the actual profile, the predictions derived from the surface mass density profile will result in an underestimation of the velocity dispersion. Panel e and f in Fig. \ref{fig:surface12} illustrate this scenario. In this context, we measure the l.o.s. velocity dispersion, i.e., the velocity dispersion along the $x$-axis. Binary systems are treated as single particles, the velocity is the c.m. velocity. Both measurements, which encompass MS stars exclusively and include all stellar types, yield consistent profiles. Therefore stars can effectively serve as tracers for the velocity dispersion. 
Moreover, there is a discernible reduction in the central l.o.s. velocity dispersion in both models. This observation aligns with the findings of \cite{2016MNRAS.458.1450W}, which demonstrated that the BH subsystem results in an incongruent outcome when fitting King's model using both the velocity dispersion profile and surface luminosity density profile. Figs. \ref{fig:surface34} and \ref{fig:surface56} demonstrate that models 3 to 6 also conform to the conclusions mentioned above. Since models 5 and 6 (fig. \ref{fig:surface56}) lack a DSC phase, the mismatch between the surface density profile and the l.o.s velocity dispersion can only be used as a tool to confirm the presence of a dark core, rather than to directly determine whether a cluster is a DSC.

\subsubsection{Mass segregation}
Gravitational heavy-star--remnant interactions lead to a heating of the stellar component, breaking down mass segregation among stars. Heavy stars are ejected from the core and can orbit at larger distances around the cluster. In this project, due to the limited number of particles (a few $10^3$ in the DSC phase or at a comparable age), measuring the radial mean mass tendency results in significant statistical uncertainty. To address this, we employ the minimum spanning tree (MST) method, as described in \cite{mst1}, to assess mass segregation.
In this method, each star in a star cluster is treated as an individual vertex, and a minimum spanning tree is constructed based on their Euclidean distances. This tree connects all vertices while minimizing the total path length. For a star cluster with $N$ stars, we select the $n$ most massive stars, and denote the path length of this subset as $l^{\rm mass}_{\rm MST}$. As a reference, we create $k$ subsets, each containing $n$ random stars from the cluster. We calculate the mean value of the path lengths $\langle l^{\rm ref}_{\rm MST} \rangle$ and their standard deviation $\Delta l^{\rm ref}_{\rm MST}$. Mass segregation is quantified using the ratios
\begin{equation}
	\label{eq:mst}
	\Lambda_{\rm MST} = \frac{\langle l^{\rm ref}_{\rm MST} \rangle}{l^{\rm mass}_{\rm MST}},\quad \Delta \Lambda_{\rm MST} = \frac{\Delta l^{\rm ref}_{\rm MST}}{l^{\rm mass}_{\rm MST}}.
\end{equation}
The second equation provides the standard deviation of $\Lambda_{\rm MST}$. In a non-segregated cluster, $\Lambda_{\rm MST}$ should be equal to 1. Conversely, if the cluster is segregated, the path length of high-mass stars should be shorter because they are more concentrated, resulting in $\Lambda_{\rm MST}$ being greater than 1. The parameters $k$ and $n$ can be chosen freely. We determine $k$ as done in \cite{mst2}, where it is selected such that a fraction $p$ of the entire $N$ stars in the cluster is covered independently of the sample size $n$
\begin{equation}
	p = 1 - \left( \frac{N-n}{N} \right)^k \rightarrow k = \lceil\frac{\mathrm{ln}(1-p)}{\mathrm{ln}(1- n/N)} \rceil,
\end{equation}
where $\lceil\cdot\rceil$ represents the ceiling function. The MST analysis requires a relatively small number of stars but can introduce large uncertainties when close binaries are present. To address this, we treat all binaries as single particles, with the mass representing the total mass of both components. The positions and velocities of the binary systems are calculated based on the center-of-mass positions and velocities.

\begin{figure*}
	\centering 
	\includegraphics[width=0.8\textwidth]{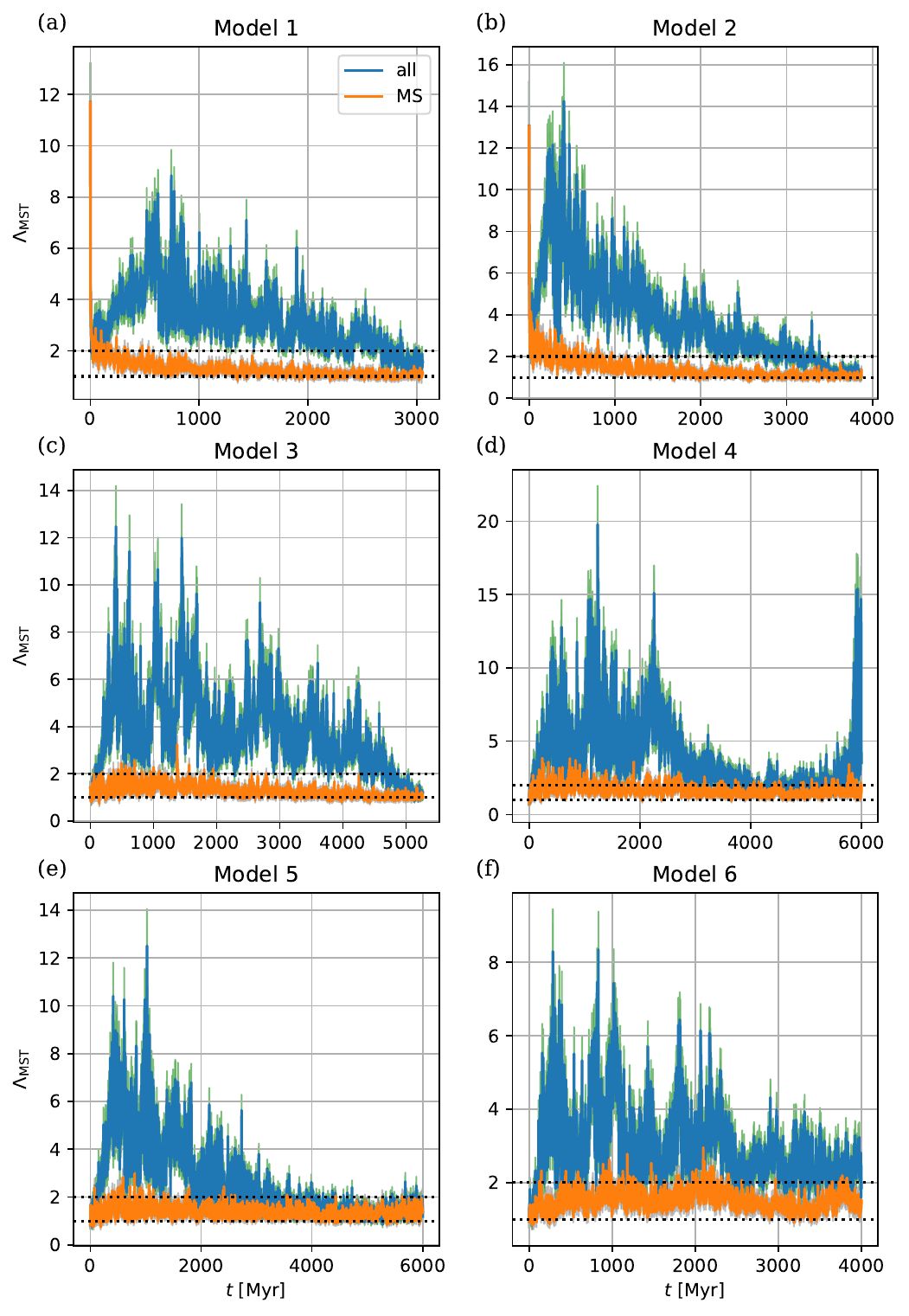}
	\caption{The measured mass segregation parameter $\Lambda_{\rm MST}$ for all models. Here we use $n=20$ and $p=0.99$. The green region denotes the 1$\sigma$ area (Not clearly visible due to the fluctuation of $\Lambda_{\rm MST}$). The horizontal dotted lines marks $\Lambda_{\rm MST}=1$ and 2. The blue lines are calculated with all stellar types and the orange lines include only MS stars.}
	\label{fig:mstn20} 
\end{figure*}

In Fig. \ref{fig:mstn20}, we present the mass segregation parameter $\Lambda_{\rm MST}$ for $n=20$ and $p=0.99$, along with its standard deviation. For model 1, 2, and 3, the actual $\Lambda_{\rm MST}$ indicates mass segregation during the DSC phase, as expected due to the presence of BHs and NSs, which are typically the most massive objects. The apparent $\Lambda_{\rm MST}$ values for model 1, 2, and 3 are consistently close to 1, suggesting no significant mass segregation or only a weak one. Previous studies, as demonstrated in Fig. 8 and 10 of \cite{clusterintidal}, have shown that star clusters typically exhibit mass segregation until their dissolving phase if there is no dark core. In model 5, where natal kicks remove most of the remnants, both $\Lambda_{\rm MST}$ values for all stellar types and MS stars only are approximately 1.5 after 3000 Gyr. Given that, at this age, the mass spectrum of the cluster occupies a narrow mass range, $\Lambda_{\rm MST} = 1.5$ already indicates a level of mass segregation in the cluster. For model 6, where binaries have higher masses than single stars, we expect an even larger $\Lambda_{\rm MST}$.

\subsubsection{Mass spectrum}
\begin{figure*}
	\centering 
	\includegraphics[width=1.6\columnwidth]{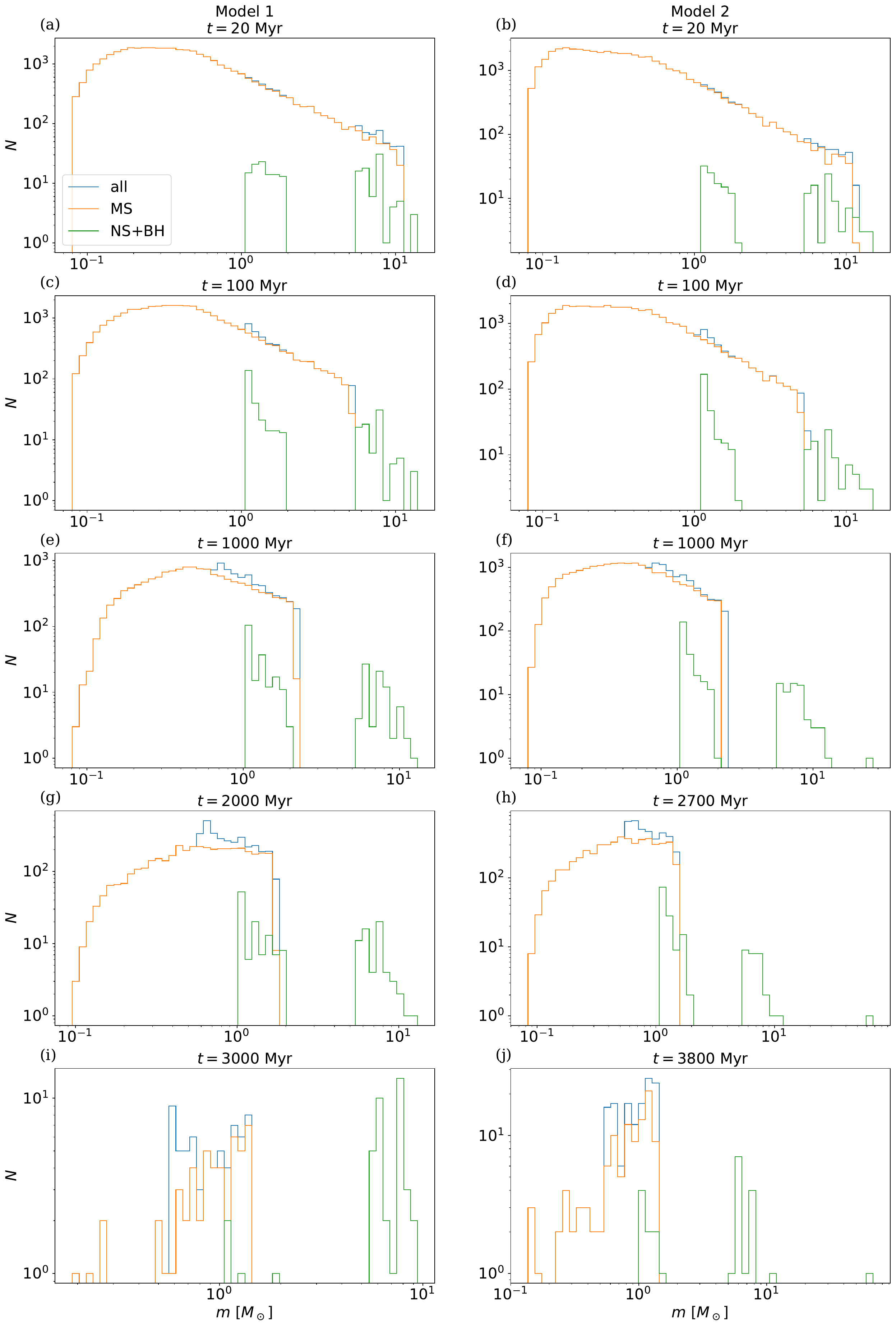}
	\caption{Mass spectrum at different times for model 1 (left panels) and model 2 (right panels). The mass spectrum is shown for all stellar types, MS stars only, and NSs and BHs separately. Panels g and h represent the DSC phase, while panels i and j correspond to a much later phase. The histograms use equal bins in mass log-space to illustrate the mass distribution in these models.}
    \label{fig:mdist12} 
\end{figure*}

The mass spectrum is displayed as a histogram in Fig. \ref{fig:mdist12} for model 1 and 2. Before entering the DSC phase, the slope of the stellar mass function continues to decrease for stars less massive than 1 $M_\odot$. However, at a very late epoch (panels i and j), the cluster is almost a two-mass-component system. In both models, the mass spectrum exhibits two prominent groups: one concentrated in a narrow range around 1 $M_\odot$, representing stars and NSs, and the other near 10 $M_\odot$, corresponding to BHs. If such a mass distribution pattern is observed, it would provide valuable insights into the dynamical properties of star clusters, assuming the existence of an invisible remnant subsystem. In model 6, we find objects with masses around 0.01 $M_\odot$. They are all WDs  and a part of a binary system (with initial masses from 2 to 9 $M_\odot$). By investigating the output from the SSE/BSE routine, we find that they are formed due to mass transfer via Roche-lobe overflow. 
\begin{figure*}
	\centering 
	\includegraphics[width=1.6\columnwidth]{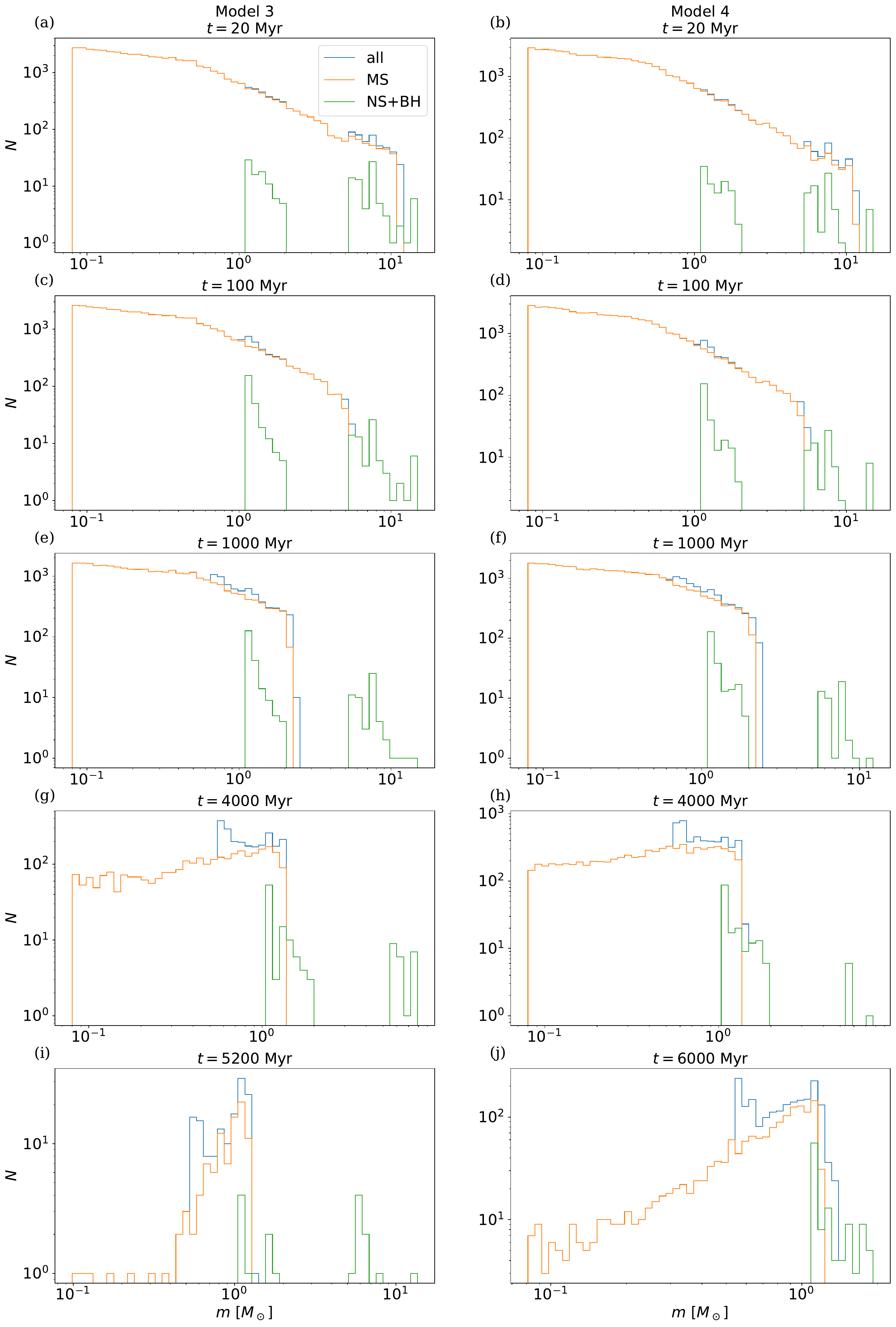}
	\caption{As Fig. \ref{fig:mdist12} but for model 3 and 4.}
    \label{fig:mdist34} 
\end{figure*}
\begin{figure*}
	\centering 
	\includegraphics[width=1.6\columnwidth]{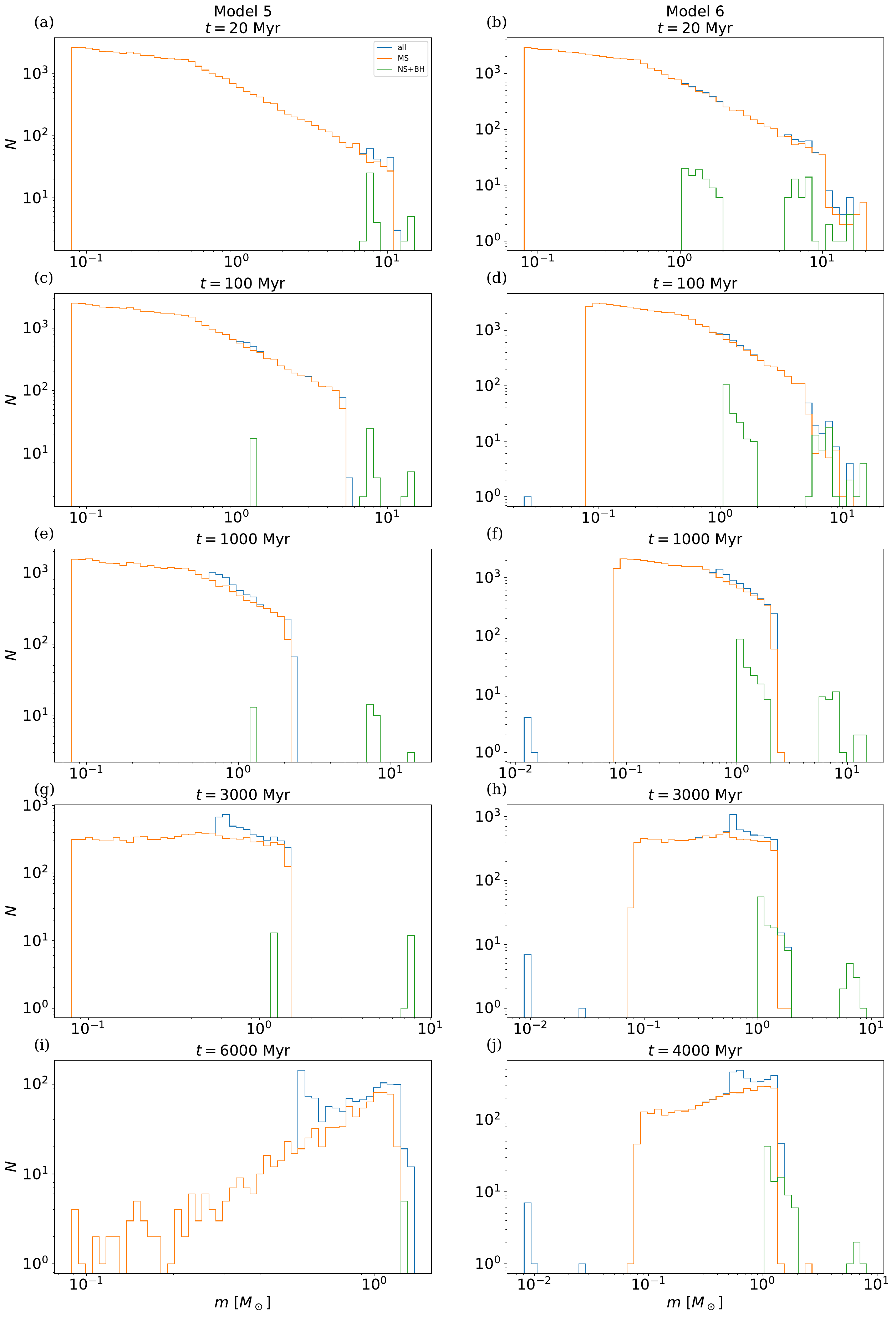}
	\caption{As Fig. \ref{fig:mdist12} but for model 5 and 6.}
    \label{fig:mdist56} 
\end{figure*}

\subsubsection{Mass to light ratio}
\begin{figure*}
	\centering 
	\includegraphics[width=\columnwidth]{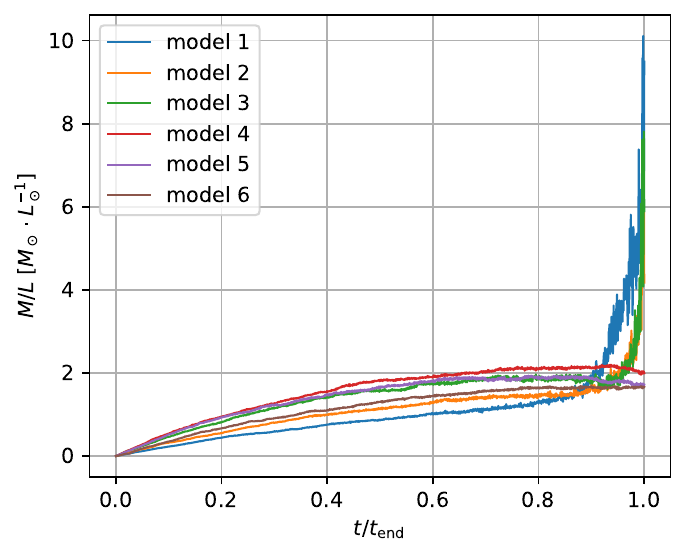}
	\caption{The evolution of the mass-to-light ratio ($M/L$) is detailed for all models. Here, the mass encompasses contributions from all stellar types, while the luminosity calculations are exclusive to MS stars. Time is normalized to the dissolution time $t_{\rm end}$, representing the maximum duration of the calculation.}
    \label{fig:mtol} 
\end{figure*}

In Figure \ref{fig:mtol}, we illustrate the evolution of the mass-to-light ratio ($M/L$). Our findings indicate that for a DSC—observed in the later stages of models 1, 2, and 3—the $M/L$ ratio escalates to approximately 10. In contrast, for the remaining models, the $M/L$ ratio stabilizes at around 2. However, employing the mass-to-light ratio as a diagnostic tool for identifying a DSC necessitates a more accurate estimation of the cluster's mass. Achieving this precision presents a considerable challenge and requires detailed measurements of the kinematics of each star.

\subsection{Initial conditions}
In this subsection, we provide a qualitative discussion of the influence of the initial half-mass radius (which affects the initial concentration of the cluster), primordial mass segregation, and the inclusion of primordial binaries.
\subsubsection{Initial half mass radius}
\label{sec:rh0}
From our observations of model 1 and model 2, it becomes evident that the formation time of a DSC depends on the initial half-mass radius. The creation of a dark core necessitates the spatial separation between the BH-NS subsystem and the luminous stars. In scenarios characterized by a smaller initial half-mass radius, as observed in Fig. \ref{fig:struct12} panels b and c, the segregation of remnants and the formation of a dark core occur more rapidly in model 2 than in model 1. This accelerated segregation and dark core formation subsequently lead to an increase in the 10\% Lagrangian radius (see panel a and b of Fig. \ref{fig:struct12}). The compact configuration of model 2 results in a lower escape rate of luminous stars, thereby delaying the transition to the DSC phase. In the case of a sufficiently small initial half-mass radius, the star cluster does not enter the DSC phase, as not enough luminous stars escape the cluster before the dissolution of the dark core.
\subsubsection{Primordial mass segregation}
The presence of primordial mass segregation plays a significant role in influencing the spatial distribution of massive remnants and luminous stars. When examining the mass spectrum of all models at 20 Myr (refer to panel a and b in Fig. \ref{fig:mdist12}, \ref{fig:mdist34}, and \ref{fig:mdist56}), it is evident that there is no distinct mass separation of Black Holes (BHs) and Neutron Stars (NSs) within the mass spectrum, when the majority of BHs are formed. For further clarification, consider panel b and c in Fig. \ref{fig:mstn20}. During this stage, the process of mass segregation is still ongoing when there is no primordial mass segregation included (indicated by the tendency of $\Lambda_{\rm MST}$ to increase). This ongoing segregation process consequently delays the formation of the dark core. As a result, the DSC entry point for model 3 is significantly later compared to model 1 and 2.
In the case of model 4, as discussed in Section \ref{sec:rh0}, the formation of the dark core is initially slowed down due to the smaller half-mass radius. However, the absence of primordial mass segregation further exacerbates this delay. Consequently, insufficient remnants remain within the cluster to form a dark core, leading to the absence of a DSC phase in this model.
\subsubsection{Primordial binary}
The presence of primordial binaries has a profound impact on the formation of a dark core that relies on mass segregation. Close binaries essentially act as single stars with higher masses within the cluster until a three-body encounter takes place. As a result, the mass segregation among single stars is disrupted, and the individual stellar types no longer play a significant role. This phenomenon is in line with the findings of \cite{2013ApJ...779...30G}, which demonstrate that binaries tend to concentrate in the later stages of a star cluster's evolution due to mass segregation.

\begin{figure*}
	\centering 
	\includegraphics[width=1.6\columnwidth]{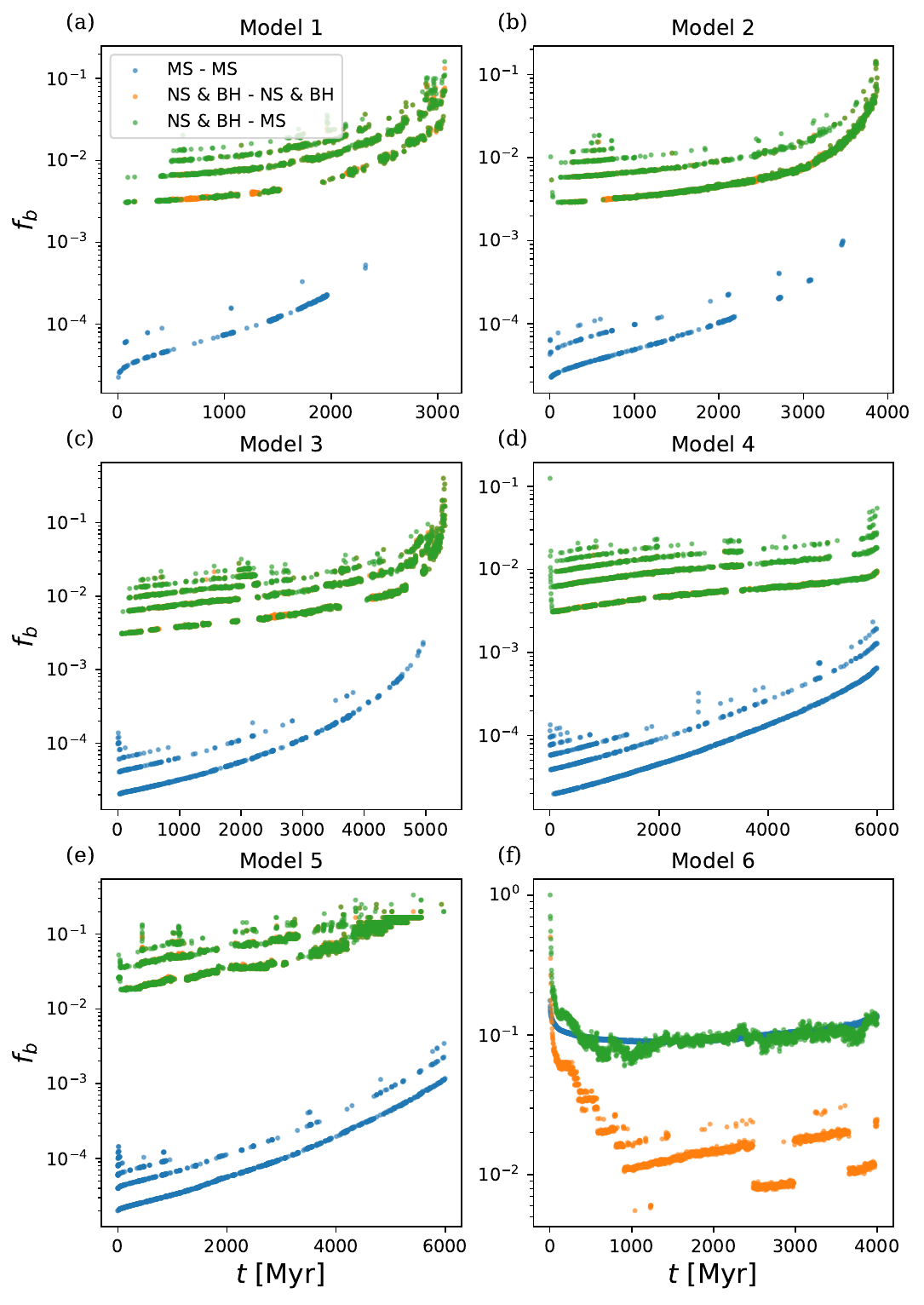}
	\caption{The evolution of the binary fraction for different stellar types in each model is depicted. In the "MS - MS" category, binary systems with two MS stars are included, and the fraction is expressed relative to the total number of MS stars. In the NS \& BH - NS \& BH category, binary systems with both components being either NS or BH are considered, and the fraction is relative to the total number of BHs and NSs. In the NS \& BH - MS category, binary systems composed of one MS star and one BH or NS are included, and the fraction is relative to the total number of BHs, NSs, and MS stars bound to them.}
    \label{fig:fbs} 
\end{figure*}

We detect binaries when the separation of two particles is shorter than 0.1 pc and the system has a negative binding energy. The binary fraction for various stellar types is illustrated in Fig. \ref{fig:fbs}. In models without primordial binaries (models 1 to 5), less than 1\% of MS stars form binaries during the cluster's evolution. However, the dark core continues to produce binaries due to three body encounters. These binaries tend to remain in the cluster for a more extended period than single BHs or NSs, leading to an increase in the binary fraction as the cluster loses mass. For all models we find that a few per-cent of BHs and NSs are in binary systems. They could be the source of GW events and the GW recoil is necessary to be considered in future works.

In our model 6, the DSC phase is not detected. However, there are still approximately 100 BHs and NSs remaining in the star cluster. From panel f of Fig. \ref{fig:fbs}, we find that about 10\% MS stars are in binaries and due to the higher mass of binaries, remnants are not able to spatially separated from luminous stars and therefore no dark core is formed.

\subsection{Lifetime}
In the study by \cite{2013MNRAS.432.2779B}, it is demonstrated that the development of a BH subsystem is intricately tied to the overall energy requirements of the entire cluster. Moreover, the decline of this subsystem is intricately regulated by the half-mass relaxation time of the cluster, which is analyzed through a semi-analytical approach considering a two-component system. In a complementary study by \cite{2020MNRAS.491.2413W}, a two-stage formation process of the BH subsystem is identified, involving mass segregation and achieving a state of balance. In this context, we evaluate the energy flux denoted as $k$ within the BH-NS subsystem.
\begin{equation}
\label{eq:k}
    \frac{E_{\rm kin}}{T^\prime_{\rm rh}} = k\frac{E_{\rm kin, sub}}{T^\prime_{\rm rh, sub}},
\end{equation}
where $E_{\rm kin}$ is the total kinetic energy of all objects within the half mass radius (calculated with all objects) of the cluster, $E_{\rm kin, sub}$ is the total kinetic energy of all objects within the half mass radius of the BH-NS subsystem but includes also luminous stars, $T^\prime_{\rm rh}$ is the corrected half mass relaxation time. We first calculate the two body half mass relaxation time
\begin{equation}
    T_{\rm rh} = 0.138 \frac{N^{1 / 2} r_{\mathrm{h}}^{3 / 2}}{\langle m\rangle^{1 / 2} G^{1 / 2} \ln \Lambda},
\end{equation}
where $N$ is the number of objects, $\langle m\rangle$ is the mean mass inside the half mass radius and $\ln \Lambda$ is the Coulomb logarithm. We use $\Lambda = 0.02N$ based on the measurement of \cite{1996MNRAS.279.1037G} for a multi-mass system. The corrected half mass relaxation time is derived by introducing a factor $\psi$
\begin{equation}
    T^\prime_{\rm rh} = \frac{T^\prime_{\rm rh}}{\psi},
\end{equation}
and
\begin{equation}
    \psi=\frac{\sum_k n_{\mathrm{k}} m_{\mathrm{k}}^2 / v_{\mathrm{k}}}{\langle n\rangle\langle m\rangle^2 /\langle v\rangle},
\end{equation}
where $n_k$, $m_k$ and $v_k$ are the number density, the mass of
one object and the mean velocity of the mass component $k$. $\langle n\rangle$,
$\langle m\rangle$ and $\langle v\rangle$ represent the average values of all components, respectively. Here we distribute all particles in 10 equal mass bins, for each $m_k$ being the mean mass of the $k$-th mass bin. The mass spread of each bin is approximately 1 $M_\odot$. Consider the mass spectrum (Fig. \ref{fig:mdist12}, \ref{fig:mdist34} and \ref{fig:mdist56}), the BHs are well separated in this distribution but the NSs are still mixed with luminous stars.  For model 5, only a few tens of BHs and NSs remain in the star cluster after 1 Gyr (see panel g in Fig. \ref{fig:fig156}), 
 and the estimation of the two-body relaxation time is unreliable for such a small number of particles. For model 6, due to the high binary fraction, three body relaxation contributes significantly to the dynamical evolution. Thus we only consider here model 1 to 4. This method is designed for two-mass components systems (one for stars, one for BHs). For multi-mass systems, i.e. as for our models, more complexity is introduced. Here we take the energy flux $k$ as a parameter for the evolution of the dark core together with other properties.

\begin{figure*}
	\centering 
	\includegraphics[width=\textwidth]{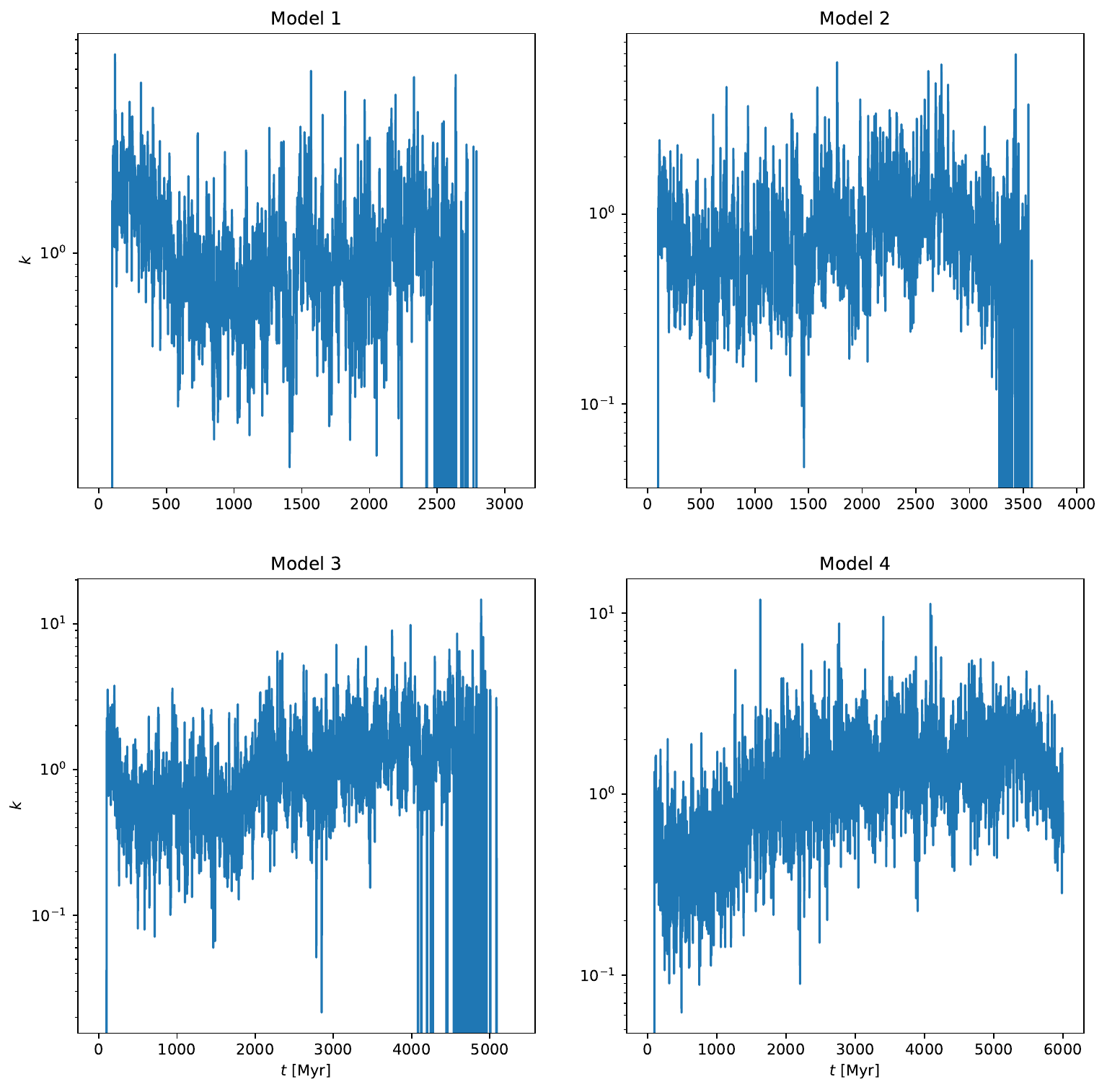}
	\caption{The energy flux $k$ is calculated for models 1, 2, 3, and 4, following the equation (\ref{eq:k}). We have distributed all particles into 10 equal mass bins, with each bin having a mass spread of approximately 1 $M_\odot$.}
	\label{fig:ks} 
\end{figure*}

Fig. \ref{fig:ks} presents the results. In the case of model 1, we can clearly observe the mass segregation phase (initial decrease), and the balance phase during which $k\approx \mathrm{constant}$ is notably shorter compared to the findings in \cite{2020MNRAS.491.2413W}. Following the balance phase, the flux $k$ begins to increase, indicating the dissolution of the dark core. As for model 2 and model 3, the balance phase is considerably shorter, resulting in a reduced lifespan for the dark core. In the case of model 4, the mass segregation phase is almost imperceptible, and no balance phase can be identified. This suggests that the dark core has an extremely brief lifespan, precluding the existence of a DSC phase in model 4.

Previous studies \citep[e.g. ][]{2015MNRAS.449L.100C,2019MNRAS.487.2412G,2021NatAs...5..957G,2023MNRAS.522.5340G}
 have demonstrated that the BH-subsystem or NSs can exert a significant influence on the lifetime of a star cluster. In our model, we observe that following the DSC phase, the cluster initiates an increased rate of stellar loss, as evident in panels g and h of Fig. \ref{fig:fig112}, as well as panel g in Fig. \ref{fig:fig134}. Conversely, for a star cluster lacking a dark core, the loss rate remains nearly constant throughout the entire simulation, as indicated in panel h of Fig. \ref{fig:fig134}, and panels g and h of Fig. \ref{fig:fig156}.
As discussed in \cite{2019MNRAS.487.2412G}, the energy generation stemming from the dark core disrupts the overall energy equilibrium of the cluster, leading to its abrupt dissolution. In our models (see Fig. \ref{fig:ks}), the energy flux originating from the dark core substantially shortens the lifetime of the star cluster.

The evolution of the dark core is largely independent of the tidal field and progresses at an individual time scale. Assuming a universal natal kick for remnants and a general mass -- half-mass radius relation at the birth of a star cluster, only two crucial conditions are pertinent to DSC formation: the initial mass of the cluster and the influence of the tidal field. This observation highlights potential DSC candidates, particularly star clusters located within a few kpc to the Galactic center and those possessing substantial mass.
Another potentially influential condition is the presence of primordial mass segregation, as it significantly reduces the time required for the dark core to form.

\section{Summary and Conclusions} \label{sec:con}
We performed {\it N}-body simulations in order to study DSCs. We have investigated the mass spectrum, the mass density profile, the velocity dispersion profile and the mass segregation parameter for a DSC. By comparison with the evolution of a star cluster without a dark core, we find that there are observational characteristics for a DSC. Firstly, the observed mass density profile is not consistent with the velocity dispersion profile. The core collapse can not be found and the velocity dispersion profile shows that there is hidden mass. Secondly, the mass spectrum and the measured mass segregation parameter, $\Lambda_{\rm MST}$, show that the cluster is not mass segregated despite being astrophysically unusual.

Both of these characteristics can be explained by the existence of non-luminous remnants in the core, i.e. the dark core. For the first characteristic, the core collapse forms a compact core and contains more remnants than bright stars. The observed velocity dispersion profile implies that a significant fraction of the mass is invisible. For the second characteristic, the heating process through the dark core yields the migration of  the heaviest stars from the core to the outer region of the cluster.  In brief, the underestimation of the velocity dispersion (or an apparent overestimation of the mass from the velocity dispersion) implies that there is a hidden mass component in the cluster. The lack of observable mass segregation implies that the hidden mass exists in the form of non-luminous massive objects located in the central region. Considering in addition the virial ratio, we suggest the following strategy to find DSCs in observations:
\begin{enumerate}
	\item Select all old and apparently unbound (i.e. $Q>1$) star clusters.
	\item Measure the mass density and velocity dispersion profile. Select clusters without core collapse and with a velocity dispersion above the prediction from the observed stellar mass density profile.
	\item Measure the mass segregation parameter $\Lambda_{\rm MST}$.
\end{enumerate}
After obtaining $\Lambda_{\rm MST}\approx 1$, we can say that the cluster is very likely to be a DSC. Besides, we may look for pulsars in the cluster to identify the existence of NSs and BHs.  We also find that the mass lose rate is not constant, the heavy star--remnant interactions in the core push out the heaviest stars from the core. Thus, during the DSC phase, the mass loss rate of the cluster is increased. Towards the end before the dissolution of the cluster, almost all low mass stars have escaped and the cluster is nearly a two mass-components system. The heaviest stars remaining in the cluster have masses in a narrow range about 1 $\mathrm{M}_{\odot}$ and the remnants (i.e. BHs) are all a few 10 $M_\odot$ heavy. 

By varying the initial half-mass radius, we have observed a significant impact of the initial concentration of the star cluster on the formation time of the dark core and the emergence of a DSC. Specifically, a higher initial concentration leads to a longer time until the DSC phase is reached, i.e. a longer formation time. Because a high concentration leads to a shorter relaxation time and therefore the energy transfer between the BH subsystem and the rest of the cluster is more efficient. We anticipate the existence of an upper limit for the initial concentration beyond which the cluster cannot enter the DSC phase before dissolution.
The introduction of primordial mass segregation has been found to accelerate the formation of the dark core. The interplay between the time scale of mass segregation and the formation of massive remnants emerges as a key condition for DSC formation. In practical terms, early violent emergence from the molecular cloud core, triggered by gas expulsion, modifies the concentration of the star cluster, typically occurring within the first 2 to 3 million years from cluster birth. Subsequent re-virialization after gas expulsion introduces additional complexity to the formation of a DSC. Another complicating factor is the presence of initial binary systems. They have the potential to challenge our theory based on the mass separation between massive remnants and luminous stars. As evidenced in our 6, the energy generation resulting from binary disruption is significant and alters the dynamical properties of the entire cluster. Additionally, natal kicks play a critical role, influencing the retention of remnants within the star cluster. It is noteworthy that all these factors are not strongly correlated with the tidal field. In our view, they exhibit universal characteristics. Future research should consider these factors, bringing theoretical models closer to reality.

Observational challenges stem from several factors. Firstly, the formation of DSCs is confined to regions with strong tidal fields, primarily in the inner Galaxy. As estimated by \cite{DSC}, this occurs within $R_G\lesssim 5$ kpc or $\gtrsim 3$ kpc from the Sun. Optical observations at these distances are problematic due to limited angular resolution and the presence of interstellar dust.
To estimate the dynamical mass of the star cluster, the measurement of the l.o.s. velocity dispersion necessitates high-resolution spectroscopic data for individual stars. Regarding the mass segregation parameter $\Lambda_{\rm MST}$, the most massive MS stars are typically the most luminous, but spectroscopic observations are still essential to identify and account for unresolved binaries. Gaia \citep{2016A&A...595A...1G} has an integrated spectrometer, but only for analyzing Ca II-triplet for the measuring radial velocity of bright stars. The James Webb Space Telescope \citep{2023PASP..135d8001R} holds promise for improved photometric observations, and specific spectroscopic observations are planned for candidate stars. The final challenge pertains to cluster member selection. A DSC is ostensibly unbound, leading to some cluster members being regarded as contaminants when applying proper motion filters. The need for a more robust selection method is evident.

\section*{Acknowledgements}
We thank the anonymous referee for constructive comments, which helped improve the methodology and discussion. 
WW thanks Long Wang for his help with the usage of \texttt{PeTar}. PK acknowledges support through the DAAD-Eastern Europe exchange grant.

\section*{Data availability}
The data used are computed with the poblicly available $N$-body code \texttt{PeTar}\footnote{\href{https://github.com/lwang-astro/PeTar}{https://github.com/lwang-astro/PeTar}} and initialization code \texttt{McLuster}\footnote{\href{https://github.com/lwang-astro/mcluster}{https://github.com/lwang-astro/mcluster}}. The simulation outputs and setups are available under request. The code for data processing and the raw data behind each plot are available under request.

%%%%%%%%%%%%%%%%%%%%%%%%%%%%%%%%%%%%%%%%%%%%%%%%%%



\bibliographystyle{mnras}
\bibliography{dsc} % if your bibtex file is called example.bib

%\begin{thebibliography}{99}
%\end{thebibliography}

%%%%%%%%%%%%%%%%%%%%%%%%%%%%%%%%%%%%%%%%%%%%%%%%%%

%%%%%%%%%%%%%%%%% APPENDICES %%%%%%%%%%%%%%%%%%%%%
\appendix
\section{Used parameters for the stellar evolution routine}
\label{sec:bseparameter}
The common parameters of the BSE/SSE routine in \texttt{PeTar} for all models are listed in Table \ref{tab:petarbseparameter}. We exercise control over the natal kicks of BHs and NSs in each model using two key parameters:
\begin{itemize}
    \item \texttt{bse-bhflag}, where a value of 0 signifies no natal kick, 1 corresponds to the same kick as NSs, and 2 represents the same kick as NSs but scaled by a fallback factor.
    \item \texttt{bse-sigma}, which denotes the velocity dispersion of the Maxwell distribution employed to generate kick velocities. Setting this parameter to 0 results in all natal kicks receiving a velocity of zero.
\end{itemize}
Table \ref{tab:petarbseparameter} presents all relevant parameters for configuring the BSE/SSE routine setup. All values, with the exception of \texttt{bse-sigma} and \texttt{bse-bhflag}, are applied uniformly across all models. Table \ref{tab:petarbseparameter1} lists these two parameters for each model.
\begin{table}
    \centering
    \begin{tabular}{cc}
    \hline
    \hline
        Parameter      &Value \\
    \hline
        \texttt{bse-alpha}      &  3 \\
        \texttt{bse-beta}       &  0.125\\
        \texttt{bse-bhwacc}     &  1.5\\
        \texttt{bse-bwind}      &  0\\
        \texttt{bse-eddfac}     &  1\\
        \texttt{bse-epsnov}     &  0.001\\
        \texttt{bse-gamma}      &  -1\\
        \texttt{bse-hewind}     &  1\\
        \texttt{bse-lambda}     &  0.5\\
        \texttt{bse-metallicity}&  0.02\\
        \texttt{bse-neta}       &  0.5\\
        \texttt{bse-pts1}       &  0.05\\
        \texttt{bse-pts2}       &  0.01\\
        \texttt{bse-pts3}       &  0.02\\
        % \texttt{bse-sigma}      &  265\\
        \texttt{bse-xi}         &  1\\
        % \texttt{bse-bhflag}     &  2\\
        \texttt{bse-ceflag}     &  0\\
        \texttt{bse-ecflag}     &  1\\
        \texttt{bse-kmech}      &  1\\
        \texttt{bse-nsflag}     &  3\\
        \texttt{bse-psflag}     &  1\\
        \texttt{bse-tflag}      &  1\\
        \texttt{bse-wdflag}     &  1\\
        \texttt{bse-sigma}      &  * \\
        \texttt{bse-bhflag}     &  * \\
        \hline
    \end{tabular}
    \caption{Relevant \textsc{SSE/BSE} parameters for simulations with \textsc{PeTar}. \texttt{bse-sigma} and \texttt{bse-bhflag} are varies depend on the setup of each model. We list the used value for these two parameters in Table \ref{tab:petarbseparameter1}.}
    \label{tab:petarbseparameter}
\end{table}

\begin{table}
    \centering
    \begin{tabular}{ccc}
    \hline
    \hline
        model & \texttt{bse-sigma} & \texttt{bse-bhflag} \\
    \hline
        1 & 0 & 0\\
        2 & 0 & 0\\
        3 & 0 & 0\\
        4 & 0 & 0\\
        5 & 265 & 2\\
        6 & 0 & 0 \\
        \hline
    \end{tabular}
    \caption{Used value of \texttt{bse-sigma} and \texttt{bse-bhflag} for each model.}
    \label{tab:petarbseparameter1}
\end{table}
%%%%%%%%%%%%%%%%%%%%%%%%%%%%%%%%%%%%%%%%%%%%%%%%%%

% Don't change these lines
\bsp	% typesetting comment
\label{lastpage}
\end{document}